\documentclass[aps,showpacs,10pt,groupedaddress]{revtex4}

\usepackage{epsfig}
\usepackage{graphicx}
\usepackage{amsmath,amssymb}

\begin{document}

\title{Electron--ion binary collisions in the presence of a magnetic field}
\author{Hrachya~B.~Nersisyan}
\affiliation{Theoretical Physics Division, Institute of Radiophysics and Electronics,
Alikhanian Brothers Str. 1, 378410 Ashtarak, Armenia}
\email{hrachya@irphe.am}

\begin{abstract}
Binary collisions between ions and electrons in an external magnetic field
are considered in second-order perturbation theory, starting from the unperturbed
helical motion of the electrons. The calculations are done with the help of an
improved BC which is uniformly valid for any strength of the magnetic field and
the second-order energy and velocity transfers are treated in the interaction in
Fourier space without specifying the interaction potential. The energy transfer
is explicitly calculated for a regularized and screened potential which is both
of finite range and less singular than the Coulomb interaction at the origin and
as the limiting cases involves the Debye (i.e., screened) and Coulomb potentials.
Two particular cases are considered in detaile: (i) Ion motion parallel to the
magnetic field with an arbitrary strength. The energy transfer involves all harmonics
of the electron cyclotron motion. (ii) The ion arbitrary motion with respect to
the strong magnetic field when the electron cyclotron radius is much smaller than
other characteristic length scales (e.g., screening length, pitch of electron helix
etc.). In the latter case the energy transfer receives two contributions which are
responsible for the electron guiding center and cyclotron orbit perturbations.
\end{abstract}

\maketitle

\section{Introduction}
\label{sec:intr}

In the presence of an external magnetic field $\mathbf{B}$ the problem
of two charged particles cannot be solved in a closed form as the relative
motion and the motion of the center of mass are coupled to each other.
Therefore no theory exists for a solution of this problem that is uniformly
valid for any strength of the magnetic field and the Coulomb force between
the particles. The energy loss of ion beams and the related processes in a
magnetized plasmas which are important in many areas of physics such as transport,
heating, magnetic confinement of thermonuclear plasmas and astrophysics are
examples of physical situations where this problem arises. Recent applications
are the cooling of heavy ion beams by electrons~\cite{sor83,pot90,mes94}
and the energy transfer for heavy-ion inertial confinement fusion (ICF) (see,
e.g.,~\cite{pro05} for an overview). The classical limit of a hydrogen or
Rydberg atom in a strong magntic field also falls in this category (see,
e.g.,~\cite{has89} and references therein) but in contrast to the free-free
transitions (scattering) the total energy is negative there.

Numerical calculations have been performed for binary collisions (BC) between
magnetized electrons~\cite{sia76,gli92} and for collisions between magnetized
electrons and ions \cite{gzwi99,zwi00,zwi02}. In general the total energy $W$
of the particles interacting in a magnetic field is conserved but the relative
and center of mass energies are not conserved separetely. In addition, the
presence of the magnetic field breaks the rotational symmetry of the system
and as a consequence only the component of the angular momentum $\mathbf{L}$
parallel to the magnetic field $L_{\parallel}$ is a constant of motion. A
different situation arises for the BC between an electron and heavy ion. As
an ion is much heavier than an electron,
its uniform motion is only weakly perturbed by collisions with the electrons
and the magnetic field. In this case $L_{\parallel}$ is not conserved but there
exists a conserved generalized energy $K$ \cite{zwi02,ner03} involving the energy
of relative motion and a magnetic term.
The seemingly simple problem of a charged particle interaction in a magnetic
field is in fact a problem of considerable complexity and the additional
degree of freedom of the cyclotron orbital motion produces a chaotic system
with two degrees (or one degree for heavy ions) of freedom~\cite{gut90,sch00,bhu02}.

In this paper we consider the BC between electrons and heavy ion treating the
interaction (Coulomb) with the ion as a perturbation to the helical motion of
the magnetized electrons. This has been done previously in first order in the
ion charge $Z$ and for an ion at rest \cite{gel97} and in up to $\mathrm{O}(Z^{2})$
for uniformly moving heavy ion~\cite{toe02,ner03}. In Ref.~\cite{toe02} three
regimes are identified, depending on the relative size of the parameters
$a$ (the cyclotron radius), $s$ (the distance of the closest approach), and
$\delta$ (the pitch of the helix). In earlier kinetic approaches
\cite{sor83,pot90,mes94} only two regimes have been distinguished:
Fast collisions for $s<a$, where the Coulomb interaction is dominant and
adiabatic collisions for $s>a$, where the magnetic field is important, as the
electron performs many gyrations during the collision with the ion. The change
$\Delta E_{i}$ of the energy of the ion has been related to the square of the
momentum transfer $\Delta p$, which has been calculated up to $\mathrm{O}(Z)$.
This is somewhat unsatisfactory, as there is another $\mathrm{O}(Z^{2})$
contribution to $\Delta E_{i}$, in which the second-order momentum transfer
enters linearly. Moreover, for applications in plasma physics (e.g., for
calculation of the ion energy loss in a magnetized plasma) one calculates
the angular avereged energy transfer which vanishes within first-order
perturbation theory due to symmetry reasons and the ion energy change receives
contribution only from higher orders~\cite{ner03}. Indeed, the transport phenomena,
etc., are of order $\mathrm{O}(Z^{2})$ in the ion charge.

In this paper we consider BC between ion and electrons in the presence of a
magnetic field within the second order perturbation theory. The present paper is a continuation
of our earlier study in Ref.~\cite{ner03} where the second-order energy transfer
is calculated with the help of an improved BC treatment which is uniformly valid
for any strength of the magnetic field and does not require the specification of
the interaction potential. The paper is organized
as follows. In Sec.~\ref{sec:s1} starting from the exact equation of motion
of two charged particles moving in a magnetic field we discuss some basic
results of the exact BC treatment for the energy and velocity transfers as well
as the energy conservation. In the following Sec.~\ref{sec:s2}, we discuss
the velocity and energy transfer during BC of magnetized electrons with ions
for arbitrary magnetic fields and strengths of the electron-ion interaction
potential. We assume that the ion mass $M$ is much larger than the electron
mass $m$. The equations of motion are solved in a perturbative manner up to
the second order in $Z$ starting from the unperturbed helical motion of the
electrons in a magnetic field. Then in Secs.~\ref{sec:s3} and \ref{sec:s4}
we turn to the explicit calculation of the ion second order energy transfer.
For further applications (e.g., in cooling of ion beams) we consider the
regularized and screened interaction potential which is both of finite range
and less singular than the Coulomb interaction at the origin and as the limiting
cases involves the Debye (i.e., screened) and Coulomb potentials. In Sec.~\ref{sec:s3}
the theory is applied to the energy transfer of an heavy ion moving parallel
to the magnetic field $\mathbf{B}$ but without any restriction on $\mathbf{B}$.
The obtained energy transfer involves all cyclotron harmonics of the electron
helical motion. The case of the strong magnetic field and arbitrary motion
of the ion with respect to $\mathbf{B}$ is derived in Sec.~\ref{sec:s4}.
It is shown that the energy transfer contains two terms which are responsible
for the electron guiding center and cyclotron motion perturbations. In Sec.~\ref{sec:disc}
the results are summed up; some formulas for the second-order treatment are
presented in the Appendices~\ref{sec:app1} and ~\ref{sec:app2}.

\section{Binary collision formulation. General treatment}
\label{sec:s1}

Below we discuss the general equations of motion for two charged particles
moving in a homogeneous magnetic field and the remaining conservation laws.
From the velocity transfer we then proceed the energy transfer of particles
during binary collision process. As shown in Ref.~\cite{ner03}, the present
treatment becomes more transparent in Fourier space.

\subsection{Relative motion and conservation laws}
\label{sec:s1.1}

We consider two point charges with masses $m_{1}$, $m_{2}$ and charges $%
q_{1}e $, $q_{2}e$, respectively, moving in a homogeneous magnetic field $%
\mathbf{B}=B\mathbf{b}$. We assume that the particles interact with the
potential $q_{1}q_{2} e\!\!\!/^{2} U(\mathbf{r})$ with $e\!\!\!/^{2}=e^{2}%
/4\pi \varepsilon _{0}$, where $\varepsilon _{0}$ is the permittivity of the
vacuum and $\mathbf{r}=\mathbf{r}_{1}-\mathbf{r}_{2}$ is the relative coordinate
of colliding particles. For
charged particles the function $U(\mathbf{r})$ can be expressed, for
instance, by the Coulomb potential, $U_{\mathrm{C}}(\mathbf{r})=1/r$. In
plasma applications the infinite range of this potential is modified by the
screening, e.g. $U_{\mathrm{D}}(\mathbf{r})=e^{-r/\lambda }/r$ with a
screening length $\lambda $ which can be chosen as the Debye screening
length $\lambda _{\mathrm{D}}$, see, for example \cite{akh75}. The quantum
uncertainty principle prevents particles from falling into the center of
these potentials. In a classical picture this can be achieved by
regularization at the origin $U_{\mathrm{R}}(\mathbf{r})=\left(
1-e^{-r/\lambdabar }\right) e^{-r/\lambda }/r$, see for example \cite%
{kel63,deu77}, where $\lambdabar $ is a parameter, which may be related to
the de Broglie wavelength.

In the presence of an external magnetic field, the Lagrangian and the
corresponding equations of particles motion cannot, in general, be separated
into parts describing the relative motion and the motion of the center of
mass with velocities $\mathbf{v}$, $\mathbf{V}_{\mathrm{cm}}$ and
coordinates $\mathbf{r}$, $\mathbf{R}_{\mathrm{cm}}$, respectively (see,
e.g., \cite{sia76}). Introducing the reduced mass $1/\mu
=1/m_{1}+1/m_{2}$ the equations of motion are 
\begin{equation}
\label{eq:a1}
{\dot{\mathbf{v}}}(t)+\Omega _{4}\left[ \mathbf{v}(t)\times \mathbf{b}\right]
=-\Omega _{3}\left[ \mathbf{V}_{\mathrm{cm}}(t)\times \mathbf{b}\right] +%
\frac{q_{1}q_{2} e\!\!\!/^{2}}{\mu }\mathbf{F}\left( \mathbf{r}%
(t)\right) ,
\end{equation}%
\begin{equation}
\label{eq:a2}
{\dot{\mathbf{V}}_{\mathrm{cm}}}(t)-\Omega _{1}\left[ \mathbf{V}_{\mathrm{cm}%
}(t)\times \mathbf{b}\right] =-\Omega _{2}\left[ \mathbf{v}(t)\times \mathbf{%
b}\right] ,
\end{equation}%
where $q_{1}q_{2} e\!\!\!/^{2}\mathbf{F}\left( \mathbf{r}%
(t)\right) $ ($\mathbf{F}=-\partial U/\partial \mathbf{r}$) is the force
exerted by the particle 2 on the particle 1. (The force which acts on the
particle 2 is $-q_{1}q_{2} e\!\!\!/^{2}\mathbf{F}\left( \mathbf{r}%
(t)\right) $). The frequencies $\Omega _{1}$, $\Omega _{2}$, $\Omega _{3}$
and $\Omega _{4}$ are expressed in terms of the cyclotron frequencies $%
\omega _{c1}=\vert q_{1}\vert eB/m_{1}$ and $\omega_{c2}=\vert q_{2}\vert eB/m_{2}$
of the particles 1 and 2, respectively, 
\begin{equation}
\label{eq:a3}
\Omega _{1}=\frac{m_{1}\varsigma _{1}\omega _{c1}+m_{2}\varsigma _{2}\omega
_{c2}}{m_{1}+m_{2}},\quad \Omega _{2}=\frac{\varsigma _{2}\omega
_{c2}-\varsigma _{1}\omega _{c1}}{m_{1}+m_{2}}\mu ,
\end{equation}%
\begin{equation}
\label{eq:a4}
\Omega _{3}=\varsigma _{2}\omega _{c2}-\varsigma _{1}\omega _{c1},\quad
\Omega _{4}=-\mu \left( \frac{\varsigma _{1}\omega _{c1}}{m_{1}}+\frac{%
\varsigma _{2}\omega _{c2}}{m_{2}}\right) .
\end{equation}%
Here $\varsigma _{\nu }=\left\vert q_{\nu }\right\vert /q_{\nu }$ with $\nu
=1,2$. From Eqs.~(\ref{eq:a1}) and (\ref{eq:a2}) follows the conservation of
total energy 
\begin{equation}
\label{eq:a5}
W=\frac{(m_{1}+m_{2})V_{\mathrm{cm}}^{2}}{2}+\frac{\mu v^{2}}{2}+
q_{1}q_{2} e\!\!\!/^{2} U(\mathbf{r})=\mathrm{const},
\end{equation}%
but the relative and center of mass energies are not conserved separately.

The coupled, nonlinear differential equations~(\ref{eq:a1}) and (\ref{eq:a2}%
) completely describe the motion of the particles. They have to be
integrated numerically for a complete set of the initial conditions for
solving the scattering problem. In the case of interaction of heavy ions ($%
m_{2}=M$, $q_{2}=Z$) with electrons ($m_{1}=m$, $q_{1}=-1$), i.e. $M\gg m$,
the equations of motion can be further simplified, since $\mu \rightarrow m$%
, $\Omega _{1},\Omega _{2}\rightarrow 0$ and $\Omega _{3},\Omega
_{4}\rightarrow \omega _{c}$ (see Eqs.~(\ref{eq:a3}) and (\ref{eq:a4})),
where $\omega _{c}=eB/m$ is the cyclotron frequency of electrons. Equation~(%
\ref{eq:a2}) leads to $\mathbf{V}_{\mathrm{cm}}\rightarrow \mathbf{v}_{i}=%
\mathrm{const}$, where $\mathbf{v}_{i}$ is the heavy ion velocity, and Eq.~(%
\ref{eq:a1}) turns into 
\begin{equation}
\label{eq:a6}
{\dot{\mathbf{v}}}(t)+\omega _{c}\left[ \mathbf{v}(t)\times \mathbf{b}\right]
=-\omega _{c}\left[ \mathbf{v}_{i}\times \mathbf{b}\right] -\frac{%
Ze\!\!\!/^{2}}{m}\mathbf{F}\left( \mathbf{r}(t)\right) .
\end{equation}%
With the help of the equation of motion~(\ref{eq:a6}) it can be easily proven
that the quantity 
\begin{equation}
\label{eq:a7}
K=\frac{mv^{2}}{2}-Ze\!\!\!/^{2}U(\mathbf{r})+m\omega _{c}\mathbf{r}\left[ 
\mathbf{v}_{i}\times \mathbf{b}\right]
\end{equation}%
is a constant of motion. In contrast to the unmagnetized case, it thus
follows that the relative energy transfer during ion-electron collision is
proportional to $\delta r_{\bot }v_{i\bot }$, where $\delta r_{\bot }$ and $%
v_{i\bot }$ are the perpendicular components of the change of relative
position and the ion velocity.

\subsection{Energy loss and velocity transfer}
\label{sec:s1.2}

The rate at which the energy of an ion in a collision with an electron at
time $t$ changes is given by 
\begin{equation}
\label{eq:a8}
\frac{dE_{i}(t)}{dt}=Ze\!\!\!/^{2}\mathbf{v}_{i}\cdot \mathbf{F}\left( 
\mathbf{r}(t)\right) ,
\end{equation}%
as $Ze\!\!\!/^{2}\mathbf{F}\left( \mathbf{r}\right) $ is the force exerted
by the electron on the ion. Integration with respect to time yields the
energy transfer itself 
\begin{equation}
\label{eq:a9}
\delta E_{i}(t)=Ze\!\!\!/^{2}\int_{-\infty }^{t}\mathbf{v}_{i}\cdot \mathbf{F%
}\left( \mathbf{r}(\tau )\right) d\tau ,
\end{equation}%
which after completion of the collision becomes 
\begin{equation}
\label{eq:a10}
\Delta E_{i}=\delta E_{i}(t\rightarrow \infty )=Ze\!\!\!/^{2}\int_{-\infty
}^{\infty }\mathbf{v}_{i}\cdot \mathbf{F}\left( \mathbf{r}(\tau )\right)
d\tau .
\end{equation}%
Here one has to integrate the equations of motion for the relative
trajectories. The limit $M\gg m$ leading to Eq.~(\ref{eq:a3}) implies that
the change in the ion energy is calculated under the assumption of a
constant ion velocity. Alternatively this energy transfer can be expressed
by the velocity transferred to the electrons during the collision. For that
purpose we substitute $\mathbf{v}_{i}=\mathbf{v}_{e}(t)-\mathbf{v}(t)$ into
Eq.~(\ref{eq:a10}) and split the electron velocity into two terms $\mathbf{v}%
_{e}(t)=\mathbf{v}_{e0}(t)+\delta \mathbf{v}(t)$, where $\mathbf{v}_{e0}(t)$
describes the helical motion in the magnetic field 
\begin{equation}
\label{eq:a11}
\dot{\mathbf{v}}_{e0}+\omega _{c}\left[ \mathbf{v}_{e0}\times \mathbf{b}%
\right] =0
\end{equation}%
and $\delta \mathbf{v}(t)$ the velocity transfer (we assume that $\delta 
\mathbf{v}(t)\rightarrow 0$ at $t\rightarrow -\infty $) due to the collision
with the ion 
\begin{equation}
\label{eq:a12}
\delta \dot{\mathbf{v}}(t)+\omega _{c}\left[ \delta \mathbf{v}(t)\times 
\mathbf{b}\right] =-\frac{Ze\!\!\!/^{2}}{m}\mathbf{F}\left( \mathbf{r}%
(t)\right) .
\end{equation}%
This yields 
\begin{equation}
\label{eq:a13}
\delta E_{i}(t)=Ze\!\!\!/^{2}\left[ \int_{-\infty }^{t}\mathbf{v}_{e0}(\tau
)\cdot \mathbf{F}\left( \mathbf{r}(\tau )\right) d\tau +\int_{-\infty
}^{t}\delta \mathbf{v}(\tau )\cdot \mathbf{F}\left( \mathbf{r}(\tau )\right)
d\tau +U\left( \mathbf{r}(t)\right) \right] .
\end{equation}%
In Eq.~(\ref{eq:a13}) we take into account that at $t\rightarrow -\infty $, $%
r(t)\rightarrow \infty $ and $U(\mathbf{r}(t))\rightarrow 0$. The time
integrals in Eq.~(\ref{eq:a13}) can be done with the help of the derivative
of the scalar product $\mathbf{v}_{e0}(t)\cdot \delta \mathbf{v}(t)$. Using
the equations of motion (\ref{eq:a11}) and (\ref{eq:a12}) we obtain 
\begin{equation}
\label{eq:a14}
\frac{d}{dt}\left[ \mathbf{v}_{e0}(t)\cdot \delta \mathbf{v}(t)\right] =-%
\frac{Ze\!\!\!/^{2}}{m}\mathbf{v}_{e0}(t)\cdot \mathbf{F}\left( \mathbf{r}%
(t)\right) ,
\end{equation}%
which yields 
\begin{equation}
\label{eq:a15}
Ze\!\!\!/^{2}\int_{-\infty }^{t}\mathbf{v}_{e0}(\tau )\cdot \mathbf{F}\left( 
\mathbf{r}(\tau )\right) d\tau =-m\mathbf{v}_{e0}(t)\cdot \delta \mathbf{v}(t).
\end{equation}%
Similarly from Eq.~(\ref{eq:a12}) 
\begin{equation}
\label{eq:a16}
\frac{d}{dt}\left[ \delta \mathbf{v}(t)\right] ^{2}=-\frac{2Ze\!\!\!/^{2}}{m}%
\delta \mathbf{v}(t)\cdot \mathbf{F}\left( \mathbf{r}(t)\right)
\end{equation}%
which yields 
\begin{equation}
\label{eq:a17}
Ze\!\!\!/^{2}\int_{-\infty }^{t}\delta \mathbf{v}(\tau )\cdot \mathbf{F}%
\left( \mathbf{r}(\tau )\right) d\tau =-\frac{m}{2}\left[ \delta \mathbf{v}%
(t)\right] ^{2}.
\end{equation}%
Thus 
\begin{equation}
\label{eq:a18}
\delta E_{i}(t)=Ze\!\!\!/^{2}U\left( \mathbf{r}(t)\right) -m\mathbf{v}%
_{e0}(t)\cdot \delta \mathbf{v}(t)-\frac{m}{2}\left[ \delta \mathbf{v}(t)%
\right] ^{2}.
\end{equation}%
The last two terms in this equation represent the change in the electron
energy due to the collision 
\begin{equation}
\label{eq:a19}
\delta E_{e}(t)=\frac{m}{2}\left\{ \left[ \mathbf{v}_{e0}(t)+\delta \mathbf{v%
}(t)\right] ^{2}-\mathbf{v}_{e0}(t)\right\} =m\mathbf{v}_{e0}(t)\cdot \delta 
\mathbf{v}(t)+\frac{m}{2}\left[ \delta \mathbf{v}(t)\right] ^{2}.
\end{equation}%
This shows energy conservation 
\begin{equation}
\label{eq:a20}
\delta E_{i}(t)+\delta E_{e}(t)-Ze\!\!\!/^{2}U(\mathbf{r}(t))=0.
\end{equation}%
As $U(\mathbf{r}(t\rightarrow \infty ))=0$ the energy change of the ion can
also be calculated from the velocity transfer $\Delta \mathbf{v}=\delta 
\mathbf{v}(t\rightarrow \infty )$ with the help of 
\begin{equation}
\label{eq:a21}
\Delta E_{i}=-\delta E_{e}(t\rightarrow \infty )=-m\left( \mathbf{v}%
_{e0}(t)\cdot \Delta \mathbf{v}+\frac{1}{2}\Delta \mathbf{v}^{2}\right) ,
\end{equation}%
this method has been adopted in \cite{toe02}. In this approach the potential 
$U(\mathbf{r})$ has to be specified at an early stage. In Sec.~\ref{sec:s2}
we will show that Eq.~(\ref{eq:a10}) allows for a more general formulation
in which the cut--off at large distances and the regularization at small
distances can be treated easily.

Until now we have considered the energy transfer of an ion. In addition
this energy transfer, $\Delta E_{i}$, can be expressed by the change of the
relative energy, $\Delta E_{r}$, and the electron momentum transfer $\Delta 
\mathbf{p}=m\Delta \mathbf{v}$. Because we are dealing with heavy ion with $%
\mathbf{v}_{i}=\mathrm{const}$, the relative velocity transfer is the same
as for the electrons, i.e. $\delta \mathbf{v}(t)$. Since the unperturbed
relative velocity is $\mathbf{v}_{0}(t)=\mathbf{v}_{e0}(t)-\mathbf{v}_{i}$
(see Eq.~(\ref{eq:a26}) below) we can establish a simple relation between
energy transfers $\Delta E_{i}$ and $\Delta E_{r}$ given by 
\begin{equation}
\label{eq:a21a}
\Delta E_{i}=-\Delta E_{r}-\mathbf{v}_{i}\cdot \Delta \mathbf{p}.
\end{equation}%
It is clear that the relative energy and momentum transfers depend only on
the relative quantities and the ion velocity $\mathbf{v}_{i}$ is not
explicitly involved in $\Delta E_{r}$ and $\Delta \mathbf{p}$. Thus, having
the ion energy transfer the other quantities can be easily extracted from
Eq.~(\ref{eq:a21a}).

\section{Perturbative treatment. General theory}
\label{sec:s2}

\subsection{Trajectory correction}
\label{sec:s2.1}

In this section we seek an approximate solution of Eq.~(\ref{eq:a12}) in
which the interaction force between the ion and electrons is considered as a
perturbation. Thus we have to look for the solution of Eq.~(\ref{eq:a12})
for the variables $\mathbf{r}$ and $\mathbf{v}$ in a perturbative manner 
\begin{equation}
\label{eq:a22}
\mathbf{r}(t)=\mathbf{r}_{0}(t)+\mathbf{r}_{1}(t)+\mathbf{r}%
_{2}(t)...,\qquad \mathbf{v}(t)=\mathbf{v}_{0}(t)+\mathbf{v}_{1}(t)+\mathbf{v%
}_{2}(t)...,
\end{equation}%
where $\mathbf{r}_{0}(t),\mathbf{v}_{0}(t)$ are the unperturbed ion-electron
relative coordinate and velocity, respectively, $\mathbf{r}_{n}(t),\mathbf{v}%
_{n}(t)\propto Z^{n}\mathbf{F}_{n-1}$ ($n=1,2,...$) are the $n$th order
perturbations of $\mathbf{r}(t)$ and $\mathbf{v}(t)$, which are proportional
to $Z^{n}$. $\mathbf{F}_{n}(t)$ is the $n$th order correction to the
ion-electron interaction force. Using the expansion (\ref{eq:a22}) for the $%
n $th order corrections $\mathbf{F}_{n}$ we obtain 
\begin{equation}
\label{eq:a23}
\mathbf{F}\left( \mathbf{r}(t)\right) =\mathbf{F}_{0}\left( \mathbf{r}%
_{0}(t)\right) +\mathbf{F}_{1}\left( \mathbf{r}_{0}(t),\mathbf{r}%
_{1}(t)\right) +...,
\end{equation}%
where 
\begin{equation}
\label{eq:a24}
\mathbf{F}_{0}\left( \mathbf{r}_{0}(t)\right) =\mathbf{F}\left( \mathbf{r}%
_{0}(t)\right) =-i\int d\mathbf{k}U(\mathbf{k})\mathbf{k}\exp \left[ i%
\mathbf{k}\cdot \mathbf{r}_{0}(t)\right] ,
\end{equation}%
\begin{equation}
\label{eq:a25}
\mathbf{F}_{1}\left( \mathbf{r}_{0}(t),\mathbf{r}_{1}(t)\right) =\left.
\left( \mathbf{r}_{1}(t)\cdot \frac{\partial }{\partial \mathbf{r}}\right) 
\mathbf{F}(\mathbf{r})\right\vert _{\mathbf{r}=\mathbf{r}_{0}(t)}=\int d%
\mathbf{k}U(\mathbf{k})\mathbf{k}\left[ \mathbf{k}\cdot \mathbf{r}_{1}(t)%
\right] \exp \left[ i\mathbf{k}\cdot \mathbf{r}_{0}(t)\right] .
\end{equation}%
In Eqs.~(\ref{eq:a24}) and (\ref{eq:a25}), we have introduced the
ion-electron interaction potential $U(\mathbf{r})$ through $\mathbf{F}(%
\mathbf{r})=-\partial U(\mathbf{r})/\partial \mathbf{r}$ and the force
corrections have been written using a Fourier transformation in space.

We start with the zero-order unperturbed helical motion of the electrons.
From Eq.~(\ref{eq:a11}) we obtain 
\begin{equation}
\label{eq:a26}
\mathbf{v}_{0}(t)=\mathbf{v}_{r}+v_{e\bot }\left\{ \mathbf{u}\cos (\omega
_{c}t)+[\mathbf{b}\times \mathbf{u}]\sin (\omega _{c}t)\right\} ,
\end{equation}%
\begin{equation}
\label{eq:a27}
\mathbf{r}_{0}(t)=\mathbf{R}_{0}+\mathbf{v}_{r}t+a\left\{ \mathbf{u}\sin
(\omega _{c}t)-[\mathbf{b}\times \mathbf{u}]\cos (\omega _{c}t)\right\} ,
\end{equation}%
where $\mathbf{u}=\left( \cos \varphi ,\sin \varphi \right) $ is the unit
vector perpendicular to the magnetic field, $v_{e\parallel }$ and $v_{e\bot
} $ (with $v_{e\bot }\geq 0$) are the electron unperturbed velocity
components parallel and perpendicular to $\mathbf{b}$, respectively, $%
\mathbf{v}_{r}=v_{e\parallel }\mathbf{b}-\mathbf{v}_{i}$ is the relative
velocity of the electron guiding center, and $a=v_{e\bot }/\omega _{c}$ is
the cyclotron radius. It should be noted that in Eqs.~(\ref{eq:a26}) and (%
\ref{eq:a27}), the variables $\mathbf{u}$ and $\mathbf{R}_{0}$ are
independent and are defined by the initial conditions.

The equation for the first-order velocity correction is given by 
\begin{equation}
\label{eq:a28}
{\dot{\mathbf{v}}}_{1}(t)+\omega _{c}\left[ \mathbf{v}_{1}(t)\times \mathbf{b%
}\right] =-\frac{Ze\!\!\!/^{2}}{m}\mathbf{F}_{0}\left( \mathbf{r}%
_{0}(t)\right)
\end{equation}%
with the solutions 
\begin{equation}
\label{eq:a29}
\mathbf{v}_{1}(t)=\frac{Ze\!\!\!/^{2}}{m}\left\{ -\mathbf{b}V_{\parallel
}(t)+\mathrm{Re}\left[ \mathbf{b}\left( \mathbf{b}\cdot \mathbf{V}_{\bot
}(t)\right) -\mathbf{V}_{\bot }(t)+i\left[ \mathbf{b}\times \mathbf{V}_{\bot
}(t)\right] \right] \right\} ,
\end{equation}%
\begin{equation}
\label{eq:a30}
\mathbf{r}_{1}(t)=\frac{Ze\!\!\!/^{2}}{m}\left\{ -\mathbf{b}P_{\parallel
}(t)+\mathrm{Re}\left[ \mathbf{b}\left( \mathbf{b}\cdot \mathbf{P}_{\bot
}(t)\right) -\mathbf{P}_{\bot }(t)+i\left[ \mathbf{b}\times \mathbf{P}_{\bot
}(t)\right] \right] \right\} ,
\end{equation}%
where we have introduced the following abbreviations 
\begin{equation}
\label{eq:a31}
V_{\parallel }(t)=\int_{-\infty }^{t}d\tau \mathbf{b}\cdot \mathbf{F}%
_{0}\left( \mathbf{r}_{0}(\tau )\right) ,\quad \mathbf{V}_{\bot
}(t)=e^{i\omega _{c}t}\int_{-\infty }^{t}d\tau e^{-i\omega _{c}\tau }\mathbf{%
F}_{0}\left( \mathbf{r}_{0}(\tau )\right) ,
\end{equation}%
\begin{equation}
\label{eq:a32}
P_{\parallel }(t)=\nu +\int_{-\infty }^{t}d\tau \left( t-\tau \right) 
\mathbf{b}\cdot \mathbf{F}_{0}\left( \mathbf{r}_{0}(\tau )\right) ,\quad 
\mathbf{P}_{\bot }(t)=\frac{1}{i\omega _{c}}\int_{-\infty }^{t}d\tau \left[
e^{i\omega _{c}(t-\tau )}-1\right] \mathbf{F}_{0}\left( \mathbf{r}_{0}(\tau
)\right)
\end{equation}%
with $\nu =\left. -tV_{\parallel }(t)\right\vert _{t\rightarrow -\infty }$
and have assumed that all corrections vanish at $t\rightarrow -\infty $. For
instance, in the unscreened Coulomb case, the interaction force $\mathbf{F}%
_{0}$ must behave as $\mathbf{F}_{0}\left( \mathbf{r}_{0}(t)\right)
\rightarrow 1/t^{2}$ for $\left\vert t\right\vert \rightarrow \infty $. Thus
from Eq.~(\ref{eq:a31}) at $t\rightarrow \infty $ we obtain $V_{\parallel
}(t)\rightarrow V_{0\parallel }=\mathrm{const}$ and $\mathbf{V}_{\bot
}(t)\rightarrow e^{i\omega _{c}t}\mathbf{V}_{0\bot }$, where $\mathbf{V}%
_{0\bot }=\mathrm{const}$. The quantities $V_{0\parallel }$ and $\mathbf{V}%
_{0\bot }$ give the first order velocity correction in Eq.~(\ref{eq:a29})
after an electron-ion collision. In this limit, we find for the first order
trajectory correction from Eqs.~(\ref{eq:a31}) and (\ref{eq:a32}) $%
P_{\parallel }(t)=V_{0\parallel }t+P_{0\parallel }$, $\mathbf{P}_{\bot
}(t)=-i\left( \mathbf{V}_{0\bot }/\omega _{c}\right) e^{i\omega _{c}t}+%
\tilde{\mathbf{P}}_{0\bot }$, where 
\begin{equation}
\label{eq:a33}
P_{0\parallel }=\nu -\int_{-\infty }^{\infty }d\tau \tau \mathbf{b}\cdot 
\mathbf{F}_{0}\left( \mathbf{r}_{0}(\tau )\right) ,\quad \tilde{\mathbf{P}}%
_{0\bot }=\frac{i}{\omega _{c}}\int_{-\infty }^{\infty }d\tau \mathbf{F}%
_{0}\left( \mathbf{r}_{0}(\tau )\right) .
\end{equation}%
For the Coulomb interaction $\nu =\mathbf{b}\cdot \mathbf{v}_{r}/v_{r}^{3}$
and $\nu =0$ for any screened interaction potential. Note that for the
Coulomb interaction the second term in the first relation of Eq.~(\ref%
{eq:a33}) tends to infinity (see, e.g., \cite{ner03}). However the
contribution of this term to the ion energy change vanishes after averaging
over impact parameters.

Substituting Eqs.~(\ref{eq:a24}) and (\ref{eq:a27}) into Eq.~(\ref{eq:a32}),
and using the Fourier series of the exponential function $e^{iz\sin \theta }$
\cite{gra80}, we obtain for an arbitrary interaction potential 
\begin{equation}
\label{eq:a34}
P_{\parallel }(t)=i\int d\mathbf{k}U(\mathbf{k})(\mathbf{k}\cdot \mathbf{b}%
)e^{i\mathbf{k}\cdot \mathbf{R}_{0}}\sum\limits_{n=-\infty }^{\infty
}e^{in\psi }J_{n}(k_{\bot }a)\frac{e^{i\zeta _{n}(\mathbf{k})t}}{\left(
\zeta _{n}(\mathbf{k})-i0\right) ^{2}},
\end{equation}%
\begin{equation}
\label{eq:a35}
\mathbf{P}_{\bot }(t)=i\int d\mathbf{k}U(\mathbf{k})\mathbf{k}e^{i\mathbf{k}%
\cdot \mathbf{R}_{0}}\sum\limits_{n=-\infty }^{\infty }e^{in\psi
}J_{n}(k_{\bot }a)\frac{e^{i\zeta _{n}(\mathbf{k})t}}{\left( \zeta _{n}(%
\mathbf{k})-i0\right) \left( \zeta _{n-1}(\mathbf{k})-i0\right) },
\end{equation}%
where $J_{n}$ are the Bessel functions of the $n$th order. Here $\zeta _{n}(%
\mathbf{k})=n\omega _{c}+\mathbf{k}\cdot \mathbf{v}_{r}$, $\psi =\varphi
-\theta $, $\tan \theta =k_{y}/k_{x}$ and $k_{\bot }$ is the component of $%
\mathbf{k}$ transverse to the magnetic field. The quantities $V_{\parallel
}(t)$ and $\mathbf{V}_{\bot }(t)$ are obtained directly from Eqs.~(\ref%
{eq:a34}) and (\ref{eq:a35}) through the relations $V_{\parallel }(t)=\dot{P}%
_{\parallel }(t)$ and $\mathbf{V}_{\bot }(t)=\dot{\mathbf{P}}_{\bot }(t)$.

It should be noted that Eqs.~(\ref{eq:a29}) and (\ref{eq:a30}) give formal
but exact solutions for the velocity and trajectory\ corrections to the
unperturbed quantities in Eqs.~(\ref{eq:a26}) and (\ref{eq:a27}) if the
first order force $\mathbf{F}_{0}\left( \mathbf{r}_{0}\right) $ in Eqs.~(\ref%
{eq:a31}) and (\ref{eq:a32}) is replaced by the exact one, $\mathbf{F}\left( 
\mathbf{r}\right) $ with exact relative coordinate $\mathbf{r}$.
Substituting Eqs.~(\ref{eq:a29}) and (\ref{eq:a30}) with exact force $%
\mathbf{F}\left( \mathbf{r}\right) $ into Eq.~(\ref{eq:a21}) we obtain an
exact relation for the ion energy transfer 
\begin{eqnarray}
\label{eq:a35a}
\Delta E_{i} &=&Ze\!\!\!/^{2}\left\{ \mathcal{V}_{\parallel }v_{e\parallel
}+v_{e\bot }\left[ \left( \mathbf{u}\cdot \mathbf{V}_{c}\right) -\mathbf{V}%
_{s}\cdot \lbrack \mathbf{u}\times \mathbf{b}]\right] \right\} \\
&+& \frac{Z^{2}e\!\!\!/^{4}}{2m}\left\{ 2\left( \mathbf{V}_{c}\cdot \left[ 
\mathbf{b}\times \mathbf{V}_{s}\right] \right) -\mathcal{V}_{\parallel }^{2}-%
\left[ \mathbf{b}\times \mathbf{V}_{c}\right] ^{2}-\left[ \mathbf{b}\times 
\mathbf{V}_{s}\right] ^{2}\right\} .  \nonumber
\end{eqnarray}%
Here 
\begin{equation}
\label{eq:a35b}
\mathcal{V}_{\parallel }=\int_{-\infty }^{\infty }d\tau \mathbf{b}\cdot 
\mathbf{F}\left( \mathbf{r}(\tau )\right) ,\quad \left\{ 
\begin{array}{c}
\mathbf{V}_{s} \\ 
\mathbf{V}_{c}%
\end{array}%
\right. =\int_{-\infty }^{\infty }d\tau \mathbf{F}\left( \mathbf{r}(\tau
)\right) \left\{ 
\begin{array}{c}
\sin \left( \omega _{c}\tau \right) \\ 
\cos \left( \omega _{c}\tau \right)%
\end{array}%
\right.
\end{equation}%
are constants. The exact energy transfer (\ref{eq:a35a}) is now expressed by
only the relative coordinate $\mathbf{r}(t)$.

\subsection{First and second order energy transfers}
\label{sec:s2.2}

The total energy change of the ion during an ion-electron collision is given
by Eqs.~(\ref{eq:a10}). Insertion of Eq.~(\ref{eq:a23}) into the general
expression (\ref{eq:a10}) yields 
\begin{equation}
\label{eq:a36}
\Delta E_{i}=\Delta E_{i}^{(1)}+\Delta E_{i}^{(2)}+...,
\end{equation}%
where 
\begin{equation}
\label{eq:a37}
\Delta E_{i}^{(1)}=Ze\!\!\!/^{2}\int_{-\infty }^{\infty }dt\mathbf{v}%
_{i}\cdot \mathbf{F}_{0}\left( \mathbf{r}_{0}(t)\right) ,\quad \Delta
E_{i}^{(2)}=Ze\!\!\!/^{2}\int_{-\infty }^{\infty }dt\mathbf{v}_{i}\cdot 
\mathbf{F}_{1}\left( \mathbf{r}_{0}(t),\mathbf{r}_{1}(t)\right)
\end{equation}%
are the first- and second order energy transfer, respectively.

\subsubsection{First order energy transfer}
\label{sec:s2.2.1}

The first-order energy transfer can be obtained by substituting Eqs.~(\ref%
{eq:a24}) and (\ref{eq:a27}) into the first one of Eqs.~(\ref{eq:a37}). This
yields 
\begin{equation}
\label{eq:a38}
\Delta E_{i}^{(1)}=-2\pi iZe\!\!\!/^{2}\int d\mathbf{k}U(\mathbf{k})\left( 
\mathbf{k}\cdot \mathbf{v}_{i}\right) e^{i\mathbf{k}\cdot \mathbf{R}%
_{0}}\sum\limits_{n=-\infty }^{\infty }e^{in\psi }J_{n}(k_{\bot }a)\delta
\left( \zeta _{n}(\mathbf{k})\right) .
\end{equation}

We now introduce the variable $\mathbf{s}=\mathbf{R}_{0\bot }^{(r)}$ which
is the component of $\mathbf{R}_{0}$ perpendicular to the relative velocity
vector $\mathbf{v}_{r}$. From Eqs.~(\ref{eq:a26}) and (\ref{eq:a27}) we can
see that $\mathbf{s}$ is the distance of closest approach for the guiding
center of the electron helical motion. The energy loss is now given by the
average of $\Delta E_{i}$ with respect to the initial phase of the electrons 
$\varphi $ and the azimuthal angle of $\mathbf{s}$. For spherically
symmetric interaction potentials ($U(\mathbf{r})=U(r)$ and $U(\mathbf{k}%
)=U(k)$) the first order energy transfer gives no contribution due to
symmetry and the ion energy change receives a contribution only from higher
orders. In fact, Eq.~(\ref{eq:a38}) for the averaged first order energy
change gives 
\begin{equation}
\label{eq:a39}
\langle \Delta E_{i}^{(1)}\rangle =-2\pi iZe\!\!\!/^{2}\int d%
\mathbf{k}U(k)\left( \mathbf{k}\cdot \mathbf{v}_{i}\right) J_{0}(\kappa
s)J_{0}(k_{\bot }a)\delta \left( \mathbf{k}\cdot \mathbf{v}_{r}\right) ,
\end{equation}%
where $\kappa ^{2}=k^{2}-\left( \mathbf{k}\cdot \mathbf{n}_{r}\right) ^{2}$
and $\mathbf{n}_{r}=\mathbf{v}_{r}/v_{r}$. As the integrand is an odd
function of $\mathbf{k}$ we have $\langle \Delta E_{i}^{(1)} \rangle =0$.

\subsubsection{Second order energy transfer}
\label{sec:s2.2.2}

Inserting Eqs.~(\ref{eq:a25}), (\ref{eq:a27}), (\ref{eq:a30}), (\ref{eq:a34}%
) and (\ref{eq:a35}) into the second equation of Eqs.~(\ref{eq:a37}) one
obtains 
\begin{eqnarray}
\label{eq:a40}
\Delta E_{i}^{(2)}(\mathbf{R}_{0},\varphi ) &=&\frac{\pi iZ^{2}e\!\!\!/^{4}}{%
m}\int d\mathbf{k}d\mathbf{k^{\prime }}U(\mathbf{k})U(\mathbf{k^{\prime }}%
)\left( \mathbf{k}\cdot \mathbf{v}_{i}\right) e^{i\left( \mathbf{k+k}%
^{\prime }\right) \cdot \mathbf{R}_{0}}   \\
&\times & \sum\limits_{n;m=-\infty }^{+\infty }e^{in\psi +im\psi ^{\prime
}}J_{n}(k_{\bot }a)J_{m}(k_{\bot }^{\prime }a)\delta \left( \zeta _{n}(%
\mathbf{k})+\zeta _{m}(\mathbf{k^{\prime }})\right) G_{m}(\mathbf{k},\mathbf{%
k^{\prime }}),  \nonumber
\end{eqnarray}%
where $\psi ^{\prime }=\varphi -\theta ^{\prime }$, and 
\begin{eqnarray}
\label{eq:a41}
G_{m}(\mathbf{k},\mathbf{k^{\prime }}) &=&\frac{2g_{1}\left( \mathbf{k},%
\mathbf{k^{\prime }}\right) }{\left( \zeta _{m}(\mathbf{k^{\prime }}%
)-i0\right) ^{2}}+\frac{g_{2}\left( \mathbf{k},\mathbf{k^{\prime }}\right)
-ig_{3}\left( \mathbf{k},\mathbf{k^{\prime }}\right) }{\left( \zeta _{m}(%
\mathbf{k^{\prime }})-i0\right) \left( \zeta _{m-1}(\mathbf{k^{\prime }}%
)-i0\right) }   \\
&+& \frac{g_{2}\left( \mathbf{k},\mathbf{k^{\prime }}\right) +ig_{3}\left( 
\mathbf{k},\mathbf{k^{\prime }}\right) }{\left( \zeta _{m}(\mathbf{k^{\prime
}})-i0\right) \left( \zeta _{m+1}(\mathbf{k^{\prime }})-i0\right) }  \nonumber
\end{eqnarray}%
with 
\begin{eqnarray}
\label{eq:a41a}
g_{1}\left( \mathbf{k},\mathbf{k^{\prime }}\right) &=&-\left( \mathbf{k}%
\cdot \mathbf{b}\right) \left( \mathbf{k^{\prime }}\cdot \mathbf{b}\right)
,\quad g_{2}\left( \mathbf{k},\mathbf{k^{\prime }}\right) =\left( \mathbf{k}%
\cdot \mathbf{b}\right) \left( \mathbf{k^{\prime }}\cdot \mathbf{b}\right)
-\left( \mathbf{k}\cdot \mathbf{k^{\prime }}\right) ,   \\
g_{3}\left( \mathbf{k},\mathbf{k^{\prime }}\right) &=&\mathbf{k}\cdot \left[ 
\mathbf{k^{\prime }}\times \mathbf{b}\right] .  \nonumber
\end{eqnarray}

Next, for the practical applications, $\Delta E_{i}^{(2)}$ is averaged with
respect to the initial phase of electrons $\varphi $ and the azimuthal angle 
$\vartheta _{\mathbf{s}}$ of the impact parameter $\mathbf{s}$. After
averaging the energy transfer $\Delta E_{i}^{(2)}$ with respect to $\varphi $
the remaining part will depend on $\delta \left( \left( \mathbf{k}+\mathbf{k}%
^{\prime }\right) \cdot \mathbf{v}_{r}\right) $, i.e. the component of $%
\mathbf{k}+\mathbf{k}^{\prime }$ along the relative velocity $\mathbf{n}_{r}$%
. Thus this $\delta $-function enforces $\mathbf{k}+\mathbf{k}^{\prime }$ to
lie in the plane transverse to $\mathbf{n}_{r}$ so that $e^{i\left( \mathbf{%
k+k}^{\prime }\right) \cdot \mathbf{R}_{0}}\delta \left( \left( \mathbf{k}+%
\mathbf{k}^{\prime }\right) \cdot \mathbf{v}_{r}\right) =e^{i\mathbf{Q}\cdot 
\mathbf{s}}\delta \left( \left( \mathbf{k}+\mathbf{k}^{\prime }\right) \cdot 
\mathbf{v}_{r}\right) $, where $\mathbf{Q}=\mathbf{k}_{\bot }^{(r)}+\mathbf{k%
}_{\bot }^{\prime (r)}$ and $\mathbf{k}_{\bot }^{(r)}$ is the component of $%
\mathbf{k}$ transverse to $\mathbf{n}_{r}$. The result of the angular
averaging reads 
\begin{eqnarray}
\label{eq:a42}
\langle \Delta E_{i}^{(2)}\rangle _{\varphi ,\vartheta _{\mathbf{s}}} &=&%
\frac{\pi iZ^{2}e\!\!\!/^{4}}{m}\int d\mathbf{k}d\mathbf{k^{\prime }}U(%
\mathbf{k})U(\mathbf{k^{\prime }})\left( \mathbf{k}\cdot \mathbf{v}%
_{i}\right) J_{0}\left( Qs\right) \delta \left( \left( \mathbf{k}+\mathbf{k}%
^{\prime }\right) \cdot \mathbf{v}_{r}\right)   \\
&\times & \sum\limits_{n=-\infty }^{\infty }\left( -1\right) ^{n}e^{in\left(
\theta -\theta ^{\prime }\right) }J_{n}(k_{\bot }a)J_{n}(k_{\bot }^{\prime
}a)G_{n}(\mathbf{k},\mathbf{k^{\prime }}).  \nonumber
\end{eqnarray}%
This series representation of the second-order energy transfer is valid for
any strength of the magnetic field. Besides the direction $\mathbf{b}$ of
the magnetic field and the direction $\mathbf{n}_{r}=\mathbf{v}_{r}/v_{r}$
of the relative velocity is singled out in the argument of the $\delta $%
-function and summand of the $n$-summation. This prevents a closed
evaluation of the energy transfer (\ref{eq:a42}). However the limiting case
of an ion motion parallel to a magnetic field of arbitrary strength and the
case of an arbitrary motion in a strong magnetic field can be treated in a
straightforward manner.

The calculation of the angular-averaged relative energy and momentum
transfers is performed by the similar method as for deriving Eq.~(\ref%
{eq:a42}). Note, however, that due to the symmetry reason the transverse
part of the momentum transfer vanishes, $\mathbf{v}_{i\bot }\cdot \langle
\Delta \mathbf{p}_{\bot }\rangle =0$, and Eq.~(\ref{eq:a21a}) is reduced to $%
\langle \Delta E_{i}^{(2)}\rangle =-\langle \Delta E_{r}^{(2)}\rangle
-v_{i\parallel }\langle \Delta p_{\parallel }^{(2)}\rangle $.

It is also usefull to integrate the $\varphi ,\vartheta _{\mathbf{s}}$%
-averaged ion energy change, $\langle \Delta E_{i}^{(2)}\rangle _{\varphi
,\vartheta _{\mathbf{s}}}$, over the impact parameters $s$ in the full 2D
space. Thus we can introduce a generalize cross section \cite%
{zwi00,ner03} through the relation 
\begin{eqnarray}
\label{eq:a43}
\sigma (\mathbf{v}_{r},\mathbf{v}_{i}) &=&2\pi \int_{0}^{\infty }\langle
\Delta E_{i}^{(2)}\rangle _{\varphi ,\vartheta _{\mathbf{s}}}sds=-\frac{%
\left( 2\pi \right) ^{4}Z^{2}e\!\!\!/^{4}}{2mv_{r}}\int d\mathbf{k}|U(%
\mathbf{k})|^{2}\left( \mathbf{k}\cdot \mathbf{v}_{i}\right)   \\
&\times & \sum\limits_{n=-\infty }^{\infty }J_{n}^{2}(k_{\bot }a)\left\{
k_{\parallel }^{2}\delta ^{\prime }\left( \zeta _{n}(\mathbf{k})\right) +%
\frac{k_{\bot }^{2}}{2\omega _{c}}\left[ \delta \left( \zeta _{n+1}(\mathbf{k%
})\right) -\delta \left( \zeta _{n-1}(\mathbf{k})\right) \right] \right\} , \nonumber
\end{eqnarray}%
where $\delta ^{\prime }(x)$ defines the derivative of the $\delta $%
-function with respect to the argument. Note that for the Coulomb
interaction $U(k)=U_{\mathrm{C}}(k)$, the full 2D integration over the $%
\mathbf{s}$-space results in a logarithmic divergence of the $\mathbf{k}$%
-integration in Eq.~(\ref{eq:a43}). This will be explicitly shown in the
next section. To cure this, we should introduce the cutoff parameters $k_{%
\mathrm{min}}$ and $k_{\mathrm{max}}$, see \cite{ner03} for details.

\section{Ion parallel motion and arbitrary magnetic field}
\label{sec:s3}

The averaged energy transfer, Eq.~(\ref{eq:a42}), can be evaluated without
further approximation for an ion motion parallel to the magnetic field and
assuming axially symmetric interaction potential, $U(\mathbf{k}%
)=U(|k_{\parallel }|,k_{\perp })$. In this case the averaged energy transfer
can be represented as the sum of all cyclotron harmonics. For the parallel
motion of the ion, $\mathbf{v}_{i\perp }=0$ and $\mathbf{v}_{r}=\left(
v_{e\parallel }-v_{i\parallel }\right) \mathbf{b}=v_{r\parallel }\mathbf{b}$%
, where $v_{i\parallel }$ and $\mathbf{v}_{i\perp }$ are the components of
ion velocity along and transverse to the magnetic field, respectively. In
general case setting $\mathbf{v}_{i\perp }=0$, we have from Eq.~(\ref{eq:a42}%
) 
\begin{eqnarray}
\label{eq:a44}
\langle \Delta E_{i\parallel }^{(2)}\rangle &=&\frac{\pi
iZ^{2}e\!\!\!/^{4}v_{i\parallel }}{m|v_{r\parallel }|}\int d\mathbf{k}d%
\mathbf{k^{\prime }}U(\mathbf{k})U(\mathbf{k^{\prime }})k_{\parallel
}J_{0}\left( |\mathbf{k}_{\bot }^{\prime }+\mathbf{k}_{\bot }|s\right)
\delta ( k_{\parallel }^{\prime }+k_{\parallel })   \\
&\times & \sum\limits_{n=-\infty }^{\infty }\left( -1\right) ^{n}e^{in\left(
\theta -\theta ^{\prime }\right) }J_{n}(k_{\bot }a)J_{n}(k_{\bot }^{\prime
}a)G_{n}(\mathbf{k},\mathbf{k^{\prime }}).  \nonumber
\end{eqnarray}%
Introducing cylindrical coordinates along $\mathbf{b}$ and using the
addition theorem for the Bessel functions $J_{0}\left( |\mathbf{k}_{\bot
}^{\prime }+\mathbf{k}_{\bot }|s\right) $ \cite{gra80} the energy transfer (%
\ref{eq:a44}) finally reads 
\begin{eqnarray}
\label{eq:a45}
\langle \Delta E_{i\parallel }^{(2)}\rangle &=&\frac{4Z^{2}e\!\!\!/^{4}v_{i%
\parallel }}{m\delta ^{2}v_{r\parallel }^{3}}\sum\limits_{n=1}^{\infty
}n^{2}\left\{ 3U_{n}\left( k_{\parallel },a,s\right) +k_{\parallel }\frac{%
\partial }{\partial k_{\parallel }}U_{n}\left( k_{\parallel },a,s\right) \right.   \\
&+& \left. \frac{\delta ^{2}}{2n}\left[ V_{n}\left( k_{\parallel },s,a\right)
-V_{n+1}\left( k_{\parallel },s,a\right) +V_{n}\left( k_{\parallel
},a,s\right) -V_{n+1}\left( k_{\parallel },a,s\right) \right] \right\}
_{k_{\parallel }=n/\delta },  \nonumber
\end{eqnarray}%
where $\delta =|v_{r\parallel }|/\omega _{c}$ is the pitch of the electron
helix, divided by $2\pi $, and 
\begin{eqnarray}
\label{eq:a46}
U_{n}\left( k_{\parallel },a,s\right) &=&\left[ \frac{\left( 2\pi \right)
^{2}}{2}\int_{0}^{\infty }U\left( k_{\parallel },k_{\perp }\right)
J_{n}\left( k_{\bot }a\right) J_{n}\left( k_{\bot }s\right) k_{\perp
}dk_{\perp }\right] ^{2},   \\
\label{eq:a47}
V_{n}\left( k_{\parallel },a,s\right) &=&\left[ \frac{\left( 2\pi \right)
^{2}}{2}\int_{0}^{\infty }U\left( k_{\parallel },k_{\perp }\right)
J_{n}\left( k_{\bot }a\right) J_{n-1}\left( k_{\bot }s\right) k_{\perp
}^{2}dk_{\perp }\right] ^{2}.
\end{eqnarray}%
We recall that for the ion parallel motion the last term in Eq.~(\ref{eq:a7}%
) vanishes, i.e. the relative energy is conserved and $\langle \Delta
E_{r}^{(2)}\rangle =0$. Therefore, the relation between ion energy and
relative momentum transfers is simplified to $\langle \Delta
E_{i}^{(2)}\rangle =-v_{i\parallel }\langle \Delta p_{\parallel
}^{(2)}\rangle $.

Now we specify the electron--ion interaction. In the following we consider
the regularized screened potential $U(\mathbf{r})=U_{\mathrm{R}}(\mathbf{r})$
introduced in Sec.~\ref{sec:s1.1} with 
\begin{equation}
\label{eq:a48}
U_{\mathrm{R}}(k_{\parallel },k_{\perp })=\frac{2}{\left( 2\pi \right) ^{2}}%
\left( \frac{1}{k_{\perp }^{2}+\kappa ^{2}}-\frac{1}{k_{\perp }^{2}+\chi ^{2}%
}\right) ,
\end{equation}%
where $\kappa ^{2}=k_{\parallel }^{2}+\lambda ^{-2}$, $\chi
^{2}=k_{\parallel }^{2}+d^{-2}$ and $d^{-1}=\lambda ^{-1}+\lambdabar ^{-1}$.
Carrying out the calculation of the $k_{\perp }$-integrals in Eqs.~(\ref%
{eq:a46}) and (\ref{eq:a47}) (see, e.g., \cite{gra80}) with the potential (%
\ref{eq:a48}) and substituting into Eq.~(\ref{eq:a45}) for the regularized
screened interaction we obtain 
\begin{eqnarray}
\label{eq:a49}
\langle \Delta E_{i\parallel }^{(2)}\rangle &=&\frac{4Z^{2}e\!\!\!/^{4}v_{i%
\parallel }}{m\delta ^{2}v_{r\parallel }^{3}}\sum\limits_{n=1}^{\infty }n^{2}%
\Bigg\{3\left[ u_{n}\left( \kappa _{n}a,\kappa _{n}s\right) -u_{n}\left(
\chi _{n}a,\chi _{n}s\right) \right] ^{2}  \nonumber \\
&+& \frac{2n^{2}}{\delta ^{2}}\left[ u_{n}\left( \kappa _{n}a,\kappa
_{n}s\right) -u_{n}\left( \chi _{n}a,\chi _{n}s\right) \right] \left[ \frac{1%
}{\kappa _{n}^{2}}T_{n}\left( \kappa _{n}a,\kappa _{n}s\right) -\frac{1}{%
\chi _{n}^{2}}T_{n}\left( \chi _{n}a,\chi _{n}s\right) \right] \\
&+& 2\delta ^{2}\left[ \kappa _{n}^{2}Q_{n}\left( \kappa _{n}a,\kappa
_{n}s\right) +\chi _{n}^{2}Q_{n}\left( \chi _{n}a,\chi _{n}s\right) -2\kappa
_{n}\chi _{n}D_{n}\left( \kappa _{n}a,\kappa _{n}s;\chi _{n}a,\chi
_{n}s\right) \right] \Bigg\}.  \nonumber
\end{eqnarray}%
Here
\begin{equation}
\label{eq:a50}
\kappa _{n}^{2}=\frac{n^{2}}{\delta ^{2}}+\frac{1}{\lambda ^{2}},\quad \chi
_{n}^{2}=\frac{n^{2}}{\delta ^{2}}+\frac{1}{d^{2}},
\end{equation}%
\begin{eqnarray}
\label{eq:a51}
u_{n}(x,y) &=&I_{n}(\xi )K_{n}(\eta ),  \nonumber \\
T_{n}(x,y) &=&\xi I_{n}^{\prime }(\xi )K_{n}(\eta )+\eta K_{n}^{\prime
}(\eta )I_{n}(\xi ),   \\
Q_{n}(x,y) &=&I_{n}(\xi )K_{n}(\eta )\left[ \frac{1}{\xi }I_{n}^{\prime
}(\xi )K_{n}(\eta )+\frac{1}{\eta }K_{n}^{\prime }(\eta )I_{n}(\xi )\right]  \nonumber
\end{eqnarray}%
with $\xi =\min (x,y)$, $\eta =\max (x,y)$, and the modified Bessel
functions $I_{n}$ and $K_{n}$, 
\begin{eqnarray}
\label{eq:a52}
D_{n}\left( x,y;X,Y\right) &=&\frac{1}{4n}\left[ s_{n-1}\left( x,y\right)
s_{n-1}\left( X,Y\right) -s_{n}\left( x,y\right) s_{n}\left( X,Y\right) \right.   \\
&+&\left. s_{n-1}\left( y,x\right) s_{n-1}\left( Y,X\right) -s_{n}\left(
y,x\right) s_{n}\left( Y,X\right) \right] ,  \nonumber
\end{eqnarray}%
\begin{equation}
\label{eq:a53}
s_{n}(x,y)=\left\{ 
\begin{array}{cc}
I_{n}\left( x\right) K_{n+1}\left( y\right) & y>x \\ 
-I_{n+1}\left( y\right) K_{n}\left( x\right) & y<x \\ 
\frac{1}{2}\left[ I_{n}\left( x\right) K_{n+1}\left( x\right) -I_{n+1}\left(
x\right) K_{n}\left( x\right) \right] & y=x%
\end{array}%
\right. .
\end{equation}%
For a study of the convergence of the series in Eq.~(\ref{eq:a49}) we note
that in all terms the modified Bessel functions $I_{n}$ carry the smaller
argument $\propto \min (a,s)$, while the $K_{n}$ depend on $\max (a,s)$. At
large harmonic numbers $n$ both the indices and the arguments of these
functions are large, and they behave as $I_{n}(n\xi )$, $K_{n}(n\eta )$.
Therefore the case $s=a$ is most critical for the convergence of (\ref%
{eq:a49}). This is intuitively clear as the gyrating electron hits the ion
on such a trajectory. This should not matter for the potential (\ref{eq:a48}%
), which has been regularized near the origin for exactly that purpose.
Since at large $n\rightarrow \infty $ and at $s=a$ the summand in Eq.~(\ref%
{eq:a49}) involves the functions $I_{n}(nx)$ and $K_{n}(nx)$ and their
derivatives the further analysis can be done using the uniform asymptotic
expansions of the modified Bessel functions \cite{gra80,abr72}. Insertion of
the expansions shows indeed that the $n$th member of the series are of the
order $\mathrm{O}(n^{-4})$, so the series converges even for $s=a$. On the
other hand, the energy transfer for the unregularized potentials $U_{\mathrm{%
C}}$ and $U_{\mathrm{D}}$ diverges for $s=a$.

For the screened but unregularized potential, i.e. in the limit $\lambdabar
\rightarrow 0$, all functions $u_{n}$, $T_{n}$, $Q_{n}$ and $D_{n}$
involving $\chi _{n}$ in their arguments tend to zero. There remains 
\begin{eqnarray}
\label{eq:a54}
\langle \Delta E_{i\parallel }^{(2)}\rangle &=&\frac{4Z^{2}e\!\!\!/^{4}v_{i%
\parallel }}{m\delta ^{2}v_{r\parallel }^{3}}\sum\limits_{n=1}^{\infty }n^{2}%
\left[ 3u_{n}^{2}\left( \kappa _{n}a,\kappa _{n}s\right) \right. \\
&+&\left. \frac{2n^{2}}{(\kappa _{n}\delta )^{2}}u_{n}\left( \kappa
_{n}a,\kappa _{n}s\right) T_{n}\left( \kappa _{n}a,\kappa _{n}s\right)
+2(\kappa _{n}\delta )^{2}Q_{n}\left( \kappa _{n}a,\kappa _{n}s\right) %
\right] .  \nonumber
\end{eqnarray}%
Now insertion of the uniform expansions of the modified Bessel functions
shows that the members of this series are independent of $n$ for $%
n\rightarrow \infty $ and $s=a$, hence the series diverges. In the Coulomb
case, i.e. for $\lambda \rightarrow \infty $, $\kappa _{n}=n/\delta $, the
resulting series are geometric and can be summed in closed form 
\begin{eqnarray}
\label{eq:a55}
\langle \Delta E_{i\parallel }^{(2)}\rangle &\simeq &\frac{%
Z^{2}e\!\!\!/^{4}v_{i\parallel }}{2m\delta ^{2}v_{r\parallel }^{3}}\frac{1}{%
\sqrt{(1+\xi ^{2})(1+\eta ^{2})}}\frac{1}{\sinh \Psi (\xi ,\eta )}   \\
&\times & \left\{ \left[ 2+\Xi _{1}(\xi ,\eta )\right] e^{-\Psi (\xi ,\eta )}+%
\frac{\Xi _{2}(\xi ,\eta )}{\sinh \Psi (\xi ,\eta )}\right\}  \nonumber
\end{eqnarray}%
with $\xi =\min \left( \frac{a}{\delta },\frac{s}{\delta }\right) $, $\eta
=\max \left( \frac{a}{\delta },\frac{s}{\delta }\right) $, $\Psi (\xi ,\eta
)=\varphi (\xi )-\varphi (\eta )$ and 
\begin{equation}
\label{eq:a56}
\varphi \left( \xi \right) =\sqrt{\xi ^{2}+1}-\ln \left( \frac{\sqrt{\xi
^{2}+1}+1}{\xi }\right) ,
\end{equation}%
\begin{eqnarray}
\label{eq:a57}
\Xi _{1}(\xi ,\eta ) &=&\frac{5(1+\xi ^{2})^{3/2}}{6\xi ^{2}(1+\eta
^{2})^{3/2}}+\frac{5(1+\eta ^{2})^{3/2}}{6\eta ^{2}(1+\xi ^{2})^{3/2}}-\frac{%
1}{3}\left( \frac{1}{\xi ^{2}}+\frac{1}{\eta ^{2}}\right)  \nonumber \\
&-&\frac{(1+\eta ^{2})^{3/2}}{2\eta ^{2}(1+\xi ^{2})^{1/2}}-\frac{(1+\xi
^{2})^{3/2}}{2\xi ^{2}(1+\eta ^{2})^{1/2}},   \\
\Xi _{2}(\xi ,\eta ) &=&\frac{(1+\xi ^{2})^{3/2}}{\xi ^{2}}-\frac{(1+\eta
^{2})^{3/2}}{\eta ^{2}}.  \nonumber
\end{eqnarray}%
For the limit $|s-a|\rightarrow 0$ the Taylor expansion of the function $%
\Psi (\xi ,\eta )$ is used. This yields 
\begin{equation}
\label{eq:a58}
\langle \Delta E_{i\parallel }^{(2)}\rangle \simeq \frac{Z^{2}e\!\!%
\!/^{4}v_{i\parallel }}{m\delta v_{r\parallel }^{3}}\frac{1}{\eta (1+\eta
^{2})^{3/2}}\frac{1}{|s-a|}
\end{equation}%
which exhibits the divergence at $s=a$. Note that in Eq.~(\ref{eq:a58}) $%
\eta =a/\delta =v_{e\bot }/|v_{r\parallel }|$ and does not depend on the
strength of the magnetic field.

For later purposes we also note the limits of Eq.~(\ref{eq:a54}) for a small
electron transversal velocity with $a\ll \delta $, 
\begin{equation}
\label{eq:a59}
\langle \Delta E_{i\parallel }^{(2)}\rangle \simeq \left( \frac{Ze\!\!\!/^{2}%
}{s}\right) ^{2}\frac{2v_{i\parallel }}{mv_{r\parallel }^{3}}\left[ \left(
\kappa _{1}s\right) K_{1}(\kappa _{1}s)\right] ^{2}\left[ 1+\frac{a^{2}}{%
\delta ^{2}}\mathfrak{F}\left( \kappa _{1}\delta ,\kappa _{1}s,\kappa
_{2}s\right) \right]
\end{equation}%
with 
\begin{equation}
\label{eq:a60}
\mathfrak{F}\left( \zeta ,\lambda ,\mu \right) =\frac{3}{2}+\frac{\zeta ^{2}%
}{2}\left[ 1+\frac{2K_{1}^{\prime }\left( \lambda \right) }{\lambda
K_{1}\left( \lambda \right) }+\left( \frac{\mu ^{2}K_{2}(\mu )}{\lambda
^{2}K_{1}(\lambda )}\right) ^{2}\right] +\frac{1}{\zeta ^{2}}\left[ 1+\frac{%
\lambda K_{1}^{\prime }\left( \lambda \right) }{K_{1}\left( \lambda \right) }%
\right] .
\end{equation}%
Because of the symmetry of Eq.~(\ref{eq:a54}) in respect to its arguments
the limit of a small impact parameter $s\ll \delta $, is given by Eqs.~(\ref%
{eq:a59}) and (\ref{eq:a60}) with the roles of $a$ and $s$ interchanged.
Similarly from Eq.~(\ref{eq:a49}) in the case of the regularized potential
and for vanishing cyclotron radius ($v_{e\bot }=0$) we obtain 
\begin{equation}
\label{eq:a59a}
\langle \Delta E_{i\parallel }^{(2)}\rangle =\left( \frac{Ze\!\!\!/^{2}}{s}%
\right) ^{2}\frac{2v_{i\parallel }}{mv_{r\parallel }^{3}}\left[ \left(
\kappa _{1}s\right) K_{1}(\kappa _{1}s)-\left( \chi _{1}s\right) K_{1}(\chi
_{1}s)\right] ^{2}.
\end{equation}%
Comparing the first term in Eq.~(\ref{eq:a59}) with Eq.~(\ref{eq:a59a}) we
conclude that the additional modified Bessel function in Eq.~(\ref{eq:a59a})
with the argument $\chi _{1}s$ guarantees the convergence of the energy
transfer at small impact parameter $s$.

For the practical applications and for general interaction potential in
Appendix~\ref{sec:app1} we also performe the $s$-integration of the second
order energy transfer, Eq.~(\ref{eq:a45}).

\section{Arbitrary ion motion in a strong magnetic field}
\label{sec:s4}

After the discussion of the energy loss for an ion moving parallel to a
magnetic field of arbitrary strength we return to the general case, where
the ion velocity has a component transverse to the field. As mentioned above
the integrations and summation in Eq.~(\ref{eq:a42}) cannot be done in this
case unless other simplifications are made. In the following we consider
strong magnetic fields and in Eq.~(\ref{eq:a42}) keep only the terms with $%
n=0,\pm 1$ and expand the remaining Bessel functions $J_{0}(k_{\bot }a)\simeq
1-(k_{\bot }a)^{2}/4$, $J_{1}(k_{\bot }a)\simeq k_{\bot }a/2$ with respect
to the cyclotron radius of electrons. Note that this approximation can be
alternatively formulated as a smallness of the electron transverse velocity $%
v_{e\bot }$. We obtain two contributions to the energy transfer, $\langle
\Delta E_{i}^{(2)}\rangle _{\mathrm{I}}$ which is independent of the
cyclotron radius and $\langle \Delta E_{i}^{(2)}\rangle _{\mathrm{II}}$
which is proportional to $a^{2}$. We split $\langle \Delta
E_{i}^{(2)}\rangle _{\mathrm{I}}$ according to 
\begin{equation}
\label{eq:a61}
\langle \Delta E_{i}^{(2)}\rangle _{\mathrm{I}}=\langle \Delta
E_{i}^{(2)}\rangle _{\mathrm{I1}}+\langle \Delta E_{i}^{(2)}\rangle _{%
\mathrm{I2}},
\end{equation}%
where 
\begin{eqnarray}
\label{eq:a62}
\langle \Delta E_{i}^{(2)}\rangle _{\mathrm{I1}} &=&\frac{\left( 2\pi
\right) ^{2}Z^{2}e\!\!\!/^{4}}{2m}\int d\mathbf{k}d\mathbf{k^{\prime }}U(%
\mathbf{k})U(\mathbf{k^{\prime }})\left( \mathbf{k}\cdot \mathbf{v}%
_{i}\right) (\mathbf{k}\cdot \mathbf{b})(\mathbf{k^{\prime }}\cdot \mathbf{b}%
)J_{0}\left( Qs\right)   \\
&\times & \delta \left( \left( \mathbf{k}+\mathbf{k}^{\prime }\right) \cdot 
\mathbf{v}_{r}\right) \delta ^{\prime }\left( \mathbf{k}\cdot \mathbf{v}%
_{r}\right) ,  \nonumber
\end{eqnarray}
\begin{eqnarray}
\label{eq:a63}
\langle \Delta E_{i}^{(2)}\rangle _{\mathrm{I2}} &=&\frac{2\pi
Z^{2}e\!\!\!/^{4}}{m\omega _{c}}\int d\mathbf{k}d\mathbf{k^{\prime }}U(%
\mathbf{k})U(\mathbf{k^{\prime }})\left( \mathbf{k}\cdot \mathbf{v}%
_{i}\right) J_{0}\left( Qs\right) \delta \left( \left( \mathbf{k}+\mathbf{k}%
^{\prime }\right) \cdot \mathbf{v}_{r}\right)   \\
&\times & \left[ g_{3}\left( \mathbf{k},\mathbf{k^{\prime }}\right) \left( 
\frac{1}{\mathbf{k}\cdot \mathbf{v}_{r}}-\frac{1}{\mathbf{k}\cdot \mathbf{v}%
_{r}-\omega _{c}}\right) +\pi g_{2}\left( \mathbf{k},\mathbf{k^{\prime }}%
\right) \delta \left( \mathbf{k}\cdot \mathbf{v}_{r}-\omega _{c}\right) %
\right] .  \nonumber
\end{eqnarray}
As in Sec.~\ref{sec:s3} we will assume axially symmetric potentials $U(%
\mathbf{k})=U(|k_{\parallel }|,k_{\perp })$. The integration is done by
using cylindrical coordinates oriented along $\mathbf{n}_{r}=\mathbf{v}%
_{r}/v_{r}$, i.e. any vector $\mathbf{C}$ will be represented as $\mathbf{C}%
=C_{\parallel }^{(r)}\mathbf{n}_{r}+\mathbf{C}_{\perp }^{(r)}$. For the
Bessel functions we use the addition theorem \cite{gra80}. The angular
integrals are trivial as they involve powers of trigonometric functions.
Note that the contribution of the term proportional to the function $g_{3}%
(\mathbf{k},\mathbf{k}')$ in Eq.~(\ref{eq:a63}) vanishes due to the antisymmetrical
behavior of this function with respect to the azymuthal angles of $\mathbf{k}%
^{(r)}_{\bot}$ and $\mathbf{k}'^{(r)}_{\bot}$. Then for the energy transfers
after straightforward calculations we obtain 
\begin{equation}
\label{eq:a64}
\langle \Delta E_{i}^{(2)}\rangle _{\mathrm{I1}}=\frac{Z^{2}e\!\!\!/^{4}v_{i%
\perp }^{2}}{mv_{r}^{6}}\left[ (v_{e\parallel
}^{2}-v_{i}^{2})T_{12}^{2}\left( s\right) +v_{e\parallel }v_{r\parallel
}T_{01}\left( s\right) T_{03}\left( s\right) \right] ,
\end{equation}
\begin{eqnarray}
\label{eq:a65}
\langle \Delta E_{i}^{(2)}\rangle _{\mathrm{I2}} &=&\frac{Z^{2}e\!\!\!/^{4}}{%
mv_{r}^{6}}\Bigg\{v_{i\bot }^{2}\left[ 2q^{2}\left( v_{e\parallel
}v_{i\parallel }-v_{i}^{2}\right) \mathcal{T}_{01}^{2}(q,s)-v_{e\parallel
}v_{r\parallel }\mathcal{T}_{01}(q,s)\mathcal{T}_{03}(q,s)\right] \\
&+& \left[ 2v_{r}^{2}(v_{e\parallel }v_{i\parallel }-v_{i}^{2})-v_{i\bot
}^{2}(v_{e\parallel }^{2}-v_{i}^{2})\right] \mathcal{T}_{12}^{2}(q,s)\Bigg\}, \nonumber
\end{eqnarray}%
where $q=\delta ^{-1}=\omega _{c}/v_{r}$, $T_{\nu \mu }\left( s\right) =%
\mathcal{T}_{\nu \mu }\left( 0,s\right) $ and 
\begin{equation}
\label{eq:a66}
\mathcal{T}_{\nu \mu }\left( q,s\right) =\frac{\left( 2\pi \right) ^{2}}{2}%
\int_{0}^{\infty }U\left( q,k_{\bot }\right) J_{\nu }\left( k_{\perp
}s\right) k_{\bot }^{\mu }dk_{\bot }.
\end{equation}
We now specify the interaction potential and explicitly calculate the
functions (\ref{eq:a66}). From Eqs.~(\ref{eq:a48}) and (\ref{eq:a66}) we
obtain in the regularized and screened case \cite{gra80} 
\begin{eqnarray}
\label{eq:a67}
\mathcal{T}_{12}^{\mathrm{R}}\left( q,s\right) &=&\kappa _{1}K_{1}\left(
\kappa _{1}s\right) -\chi _{1}K_{1}\left( \chi _{1}s\right) ,  \nonumber \\
\mathcal{T}_{03}^{\mathrm{R}}\left( q,s\right) &=&\chi _{1}^{2}K_{0}\left(
\chi _{1}s\right) -\kappa _{1}^{2}K_{0}\left( \kappa _{1}s\right) , \\
\mathcal{T}_{01}^{\mathrm{R}}\left( q,s\right) &=&K_{0}\left( \kappa
_{1}s\right) -K_{0}\left( \chi _{1}s\right)  \nonumber
\end{eqnarray}%
with $\kappa _{1}^{2}=q^{2}+\lambda ^{-2}$ and $\chi _{1}^{2}=q^{2}+d^{-2}$.
The functions $T_{12}^{\mathrm{R}}\left( s\right) $, $T_{03}^{\mathrm{R}%
}\left( s\right) $ and $T_{01}^{\mathrm{R}}\left( s\right) $ are easily
obtained from Eq.~(\ref{eq:a67}) setting there $q=0$, i.e. $\kappa
_{1}=1/\lambda $ and $\chi _{1}=1/d$. We investigate the asymptotic behavior
of the functions in Eq.~(\ref{eq:a67}) in the limit $s\rightarrow 0$. As $%
K_{0}\left( z\right) \sim \ln (1/z)$ and $K_{1}\left( z\right) \sim 1/z$ the
divergence is not worse than logarithmic and will cause no harm when
integrating over the impact parameter $s$. This is not so in the case of the
screened potential. Insertion of Eq.~(\ref{eq:a67}) at $\lambdabar
\rightarrow 0$ (i.e. $\chi_{1}\rightarrow \infty$) into Eq.~(\ref{eq:a64}) yields 
\begin{equation}
\label{eq:a68}
\langle \Delta E_{i}^{(2)}\rangle _{\mathrm{I1}}=\left( \frac{Ze\!\!\!/^{2}}{%
s}\right) ^{2}\frac{v_{i\perp }^{2}}{mv_{r}^{6}}\left\{ (v_{e\parallel
}^{2}-v_{i}^{2})\left[ \rho K_{1}(\rho )\right] ^{2}-v_{e\parallel
}v_{r\parallel }\left[ \rho K_{0}(\rho )\right] ^{2}\right\}
\end{equation}%
with $\rho =s/\lambda $, which behaves like $s^{-2}$ for small impact
parameters. On the other hand all functions in Eq.~(\ref{eq:a67}) vanish
exponentially for large impact parameters because of the finite range of the
potentials $U_{\mathrm{R}}$ and $U_{\mathrm{D}}$. We obtain the Coulomb case
by taking the limit $\lambda \rightarrow \infty $ in Eq.~(\ref{eq:a68}).
This yields 
\begin{equation}
\label{eq:a69}
\langle \Delta E_{i}^{(2)}\rangle _{\mathrm{I1}}=\left( \frac{Ze\!\!\!/^{2}}{%
s}\right) ^{2}\frac{v_{i\perp }^{2}}{mv_{r}^{6}}(v_{e\parallel
}^{2}-v_{i}^{2})
\end{equation}%
which is precisely the energy transfer for tight helices \cite{toe02}.

We turn now to the next term $\langle \Delta E_{i}^{(2)}\rangle _{\mathrm{I2}%
}$ in Eq.~(\ref{eq:a61}), i.e. Eq.~(\ref{eq:a65}) with functions from Eq.~(%
\ref{eq:a67}). For the screened potential $U_{\mathrm{D}}$ the insertion of
Eq.~(\ref{eq:a67}) at $\lambdabar \rightarrow 0$ yields 
\begin{eqnarray}
\label{eq:a70}
\langle \Delta E_{i}^{(2)}\rangle _{\mathrm{I2}} &=&\left( \frac{%
Ze\!\!\!/^{2}}{s}\right) ^{2}\frac{1}{mv_{r}^{6}}\Bigg\{v_{i\bot }^{2}\left[ 
\frac{2}{\left( \kappa _{1}\delta \right) ^{2}}\left( v_{e\parallel
}v_{i\parallel }-v_{i}^{2}\right) +v_{e\parallel }v_{r\parallel }\right] %
\left[ \rho _{1}K_{0}\left( \rho _{1}\right) \right] ^{2}   \\
&+& \left[ 2v_{r}^{2}(v_{e\parallel }v_{i\parallel }-v_{i}^{2})-v_{i\bot
}^{2}(v_{e\parallel }^{2}-v_{i}^{2})\right] \left[ \rho _{1}K_{1}\left( \rho
_{1}\right) \right] ^{2}\Bigg\}  \nonumber
\end{eqnarray}%
with $\rho _{1}=\kappa _{1}s$. For parallel ion motion, $v_{i\perp
}\rightarrow 0$, the first term vanishes $\langle \Delta E_{i}^{(2)}\rangle
_{\mathrm{I1}}=0$ and there remains a contribution to $\langle \Delta
E_{i}^{(2)}\rangle _{\mathrm{I2}}$, which as expected is equal to the
leading term of the corresponding expansion of Eq.~(\ref{eq:a59}) in orders
of $a/\delta $.

Taking now the limit $\lambda \rightarrow \infty $ for the unscreened
Coulomb interaction we see that $\langle \Delta E_{i}^{(2)}\rangle _{\mathrm{%
I2}}$ vanishes exponentially for $s>\delta $. Hence the energy transfer is
given by the tight helix term Eq.~(\ref{eq:a69}). For the case $s<\delta $
and hence $\rho _{1}\simeq 0$ we obtain 
\begin{equation}
\label{eq:a71}
\langle \Delta E_{i}^{(2)}\rangle _{\mathrm{I1}}+\langle \Delta
E_{i}^{(2)}\rangle _{\mathrm{I2}}=\left( \frac{Ze\!\!\!/^{2}}{s}\right) ^{2}%
\frac{2\mathbf{v}_{i}\cdot \mathbf{v}_{r}}{mv_{r}^{4}}.
\end{equation}%
This is just the stretched helix case considered in Ref.~\cite{toe02}.
Equation~(\ref{eq:a71}) is the same as the second-order energy transfer in a
field-free case (see, e.g., Refs.~\cite{ner03}) but the full relative
velocity $\mathbf{u}_{r}=\mathbf{v}_{e}-\mathbf{v}_{i}$ is replaced here by
the relative velocity of the electron guiding center $\mathbf{v}_{r}$.

These results are obtained in the limit $a\rightarrow 0$ where the electrons
move along their guiding center trajectories. Moreover, for $\omega
_{c}\rightarrow \infty $ also the pitch $\delta \rightarrow 0$ and these
trajectories are rectilinear along the lines of the magnetic field and the
energy transfer is given by $\langle \Delta E_{i}^{(2)}\rangle _{\mathrm{I1}%
} $. For a finite $\omega _{c}$ corresponding to a finite pitch the
contribution $\langle \Delta E_{i}^{(2)}\rangle _{\mathrm{I2}}$ describes
the perturbation of the guiding center trajectory.

The quadratic term $\langle \Delta E_{i}^{(2)}\rangle _{\mathrm{II}}\sim
a^{2}$ accounts for the finite cyclotron motion of the electrons. In general
this term is obtained from Eq.~(\ref{eq:a42}) and reads 
\begin{eqnarray}
\label{eq:a72}
\langle \Delta E_{i}^{(2)}\rangle _{\mathrm{II}} &=&-\frac{\pi
iZ^{2}e\!\!\!/^{4}}{4m}a^{2}\int d\mathbf{k}d\mathbf{k^{\prime }}U(\mathbf{k}%
)U(\mathbf{k^{\prime }})\left( \mathbf{k}\cdot \mathbf{v}_{i}\right)
J_{0}\left( Qs\right) \delta \left( \left( \mathbf{k}+\mathbf{k}^{\prime
}\right) \cdot \mathbf{v}_{r}\right)   \\
&\times & \left\{ \left( k_{\bot }^{2}+k_{\bot }^{\prime 2}\right) G_{0}(%
\mathbf{k},\mathbf{k^{\prime }})+k_{\bot }k_{\bot }^{\prime }\left[
e^{i\left( \theta -\theta ^{\prime }\right) }G_{1}(\mathbf{k},\mathbf{%
k^{\prime }})+e^{-i\left( \theta -\theta ^{\prime }\right) }G_{-1}(\mathbf{k}%
,\mathbf{k^{\prime }})\right] \right\} .  \nonumber
\end{eqnarray}%
Here $G_{0}(\mathbf{k},\mathbf{k^{\prime }})$ and $G_{\pm 1}(\mathbf{k},%
\mathbf{k^{\prime }})$ are given by Eq.~(\ref{eq:a41}). Using the same
techniques as before the straightforward calculation yields
\begin{equation}
\label{eq:a73}
\langle \Delta E_{i}^{(2)}\rangle _{\mathrm{II}}=\left( \frac{Ze\!\!\!/^{2}a%
}{s^{2}}\right) ^{2}\frac{1}{2mv_{r}^{2}}\left( \mathcal{Q}_{0}+\mathcal{Q}%
_{1}+\mathcal{Q}_{2}\right) .
\end{equation}%
Here we restrict ourselves to give the result for the screened potential $U_{%
\mathrm{D}}$
\begin{eqnarray}
\label{eq:a74}
\mathcal{Q}_{0} &=&\frac{v_{e\parallel }v_{r\parallel }v_{i\bot }^{2}}{%
v_{r}^{4}}\left( \frac{v_{i\bot }^{2}}{8v_{r}^{2}}-\frac{v_{r\parallel }^{2}%
}{v_{r}^{2}}\right) \left[ \rho ^{2}K_{0}\left( \rho \right) \right] ^{2}+%
\frac{v_{i\bot }^{2}}{v_{r}^{2}}\left[ \frac{v_{e\parallel }v_{i\parallel
}-v_{i}^{2}}{v_{r}^{2}}\left( 1-\frac{3v_{i\bot }^{2}}{4v_{r}^{2}}\right) \right.   \\
&+&\left. \frac{v_{e\parallel }v_{r\parallel }}{v_{r}^{2}}\left( 1-\frac{%
3v_{i\bot }^{2}}{2v_{r}^{2}}\right) \right] \left[ \rho ^{2}K_{1}\left( \rho
\right) \right] ^{2}+\frac{3v_{e\parallel }v_{r\parallel }v_{i\bot }^{4}}{%
8v_{r}^{6}}\left[ \rho ^{2}K_{2}\left( \rho \right) \right] ^{2},  \nonumber
\end{eqnarray}%
\begin{eqnarray}
\label{eq:a75}
\mathcal{Q}_{1} &=&\Bigg\{ \frac{v_{e\parallel }v_{r\parallel }v_{i\bot }^{2}%
}{v_{r}^{4}}\left( 1-\frac{17v_{i\bot }^{2}}{8v_{r}^{2}}\right) +\frac{%
v_{r\parallel }^{2}\left( v_{e\parallel }v_{i\parallel }-v_{i}^{2}\right) }{%
v_{r}^{4}}\left( 1+\frac{v_{i\bot }^{2}}{2v_{r}^{2}}\right)   \nonumber \\
&+&  \frac{q^{2}}{\kappa _{1}^{2}}\frac{v_{i\bot }^{2}}{v_{r}^{2}}\left[
\frac{v_{e\parallel }v_{r\parallel }}{v_{r}^{2}}\left( 5-\frac{12v_{i\bot
}^{2}}{v_{r}^{2}}\right) +\frac{v_{e\parallel }v_{i\parallel }-v_{i}^{2}}{%
v_{r}^{2}}\left( 10-\frac{11v_{i\bot }^{2}}{v_{r}^{2}}\right) \right]  \nonumber \\
&+& \frac{2q^{4}}{\kappa _{1}^{4}}\frac{v_{i\bot }^{2}}{v_{r}^{2}}%
\left[ \frac{v_{e\parallel }v_{r\parallel }}{v_{r}^{2}}\left( 1-\frac{%
2v_{i\bot }^{2}}{v_{r}^{2}}\right) +\frac{v_{e\parallel }v_{i\parallel
}-v_{i}^{2}}{v_{r}^{2}}\left( 7-\frac{8v_{i\bot }^{2}}{v_{r}^{2}}\right) %
\right] \Bigg\} \left[ \rho _{1}^{2}K_{0}(\rho _{1})\right] ^{2}  \nonumber \\
&+& \Bigg\{ -\frac{3v_{e\parallel }v_{r\parallel }^{3}v_{i\bot }^{2}}{%
v_{r}^{6}}+\frac{v_{e\parallel }v_{i\parallel }-v_{i}^{2}}{v_{r}^{2}}\left( 
\frac{3v_{i\bot }^{4}}{4v_{r}^{4}}+\frac{2v_{r\parallel }^{2}}{v_{r}^{2}} \right)    \\
&+&  \frac{q^{2}}{\kappa _{1}^{2}}\left[ \frac{v_{e\parallel
}v_{r\parallel }v_{i\bot }^{2}}{v_{r}^{4}}\left( \frac{13v_{i\bot }^{2}}{%
v_{r}^{2}}-7\right) +\frac{v_{e\parallel }v_{i\parallel }-v_{i}^{2}}{%
v_{r}^{2}}\left( \frac{16v_{i\bot }^{4}}{v_{r}^{4}}-\frac{23v_{i\bot }^{2}}{%
v_{r}^{2}}+6\right) \right] \Bigg\} \left[ \rho _{1}^{2}K_{1}(\rho _{1})%
\right] ^{2}  \nonumber \\
&+& \left[ \frac{v_{e\parallel }v_{r\parallel }v_{i\bot }^{2}}{2v_{r}^{4}}%
\left( 4-\frac{7v_{i\bot }^{2}}{4v_{r}^{2}}\right) -\frac{v_{e\parallel
}v_{i\parallel }-v_{i}^{2}}{v_{r}^{2}}\left( 1-\frac{3v_{i\bot }^{2}}{%
2v_{r}^{2}}+\frac{v_{i\bot }^{4}}{4v_{r}^{4}}\right) \right] \left[ \rho
_{1}^{2}K_{2}(\rho _{1})\right] ^{2}  \nonumber \\
&-& \frac{q^{2}}{\kappa _{1}^{2}}\frac{v_{e\parallel }v_{i\parallel
}-v_{i}^{2}}{v_{r}^{2}}\left[ \frac{v_{i\bot }^{2}}{v_{r}^{2}}\left( 2-\frac{%
3v_{i\bot }^{2}}{2v_{r}^{2}}\right) +\frac{2q^{2}}{\kappa _{1}^{2}}\left( 2-%
\frac{3v_{i\bot }^{2}}{v_{r}^{2}}+\frac{2v_{i\bot }^{4}}{v_{r}^{4}}\right) +%
\frac{4q^{4}}{\kappa _{1}^{4}}\frac{v_{r\parallel }^{2}v_{i\bot }^{2}}{%
v_{r}^{4}}\right] \rho _{1}^{5}K_{0}(\rho _{1})K_{1}(\rho _{1}),  \nonumber
\end{eqnarray}%
\begin{eqnarray}
\label{eq:a76}
\mathcal{Q}_{2} &=&\frac{v_{i\bot }^{4}}{v_{r}^{4}}\left( 1+\frac{8q^{2}}{%
\kappa _{2}^{2}}\right) \left[ \frac{v_{e\parallel }v_{r\parallel }}{%
v_{r}^{2}}+\frac{v_{e\parallel }v_{i\parallel }-v_{i}^{2}}{2v_{r}^{2}}\left(
1+\frac{8q^{2}}{\kappa _{2}^{2}}\right) \right] \left[ \rho
_{2}^{2}K_{0}(\rho _{2})\right] ^{2}+\frac{2v_{i\bot }^{2}}{v_{r}^{2}}%
\Bigg\{ \frac{v_{e\parallel }v_{r\parallel }}{v_{r}^{2}}   \nonumber \\
&\times &  \left( 1-\frac{3v_{i\bot }^{2}}{4v_{r}^{2}}\right) +\frac{%
4q^{2}}{\kappa _{2}^{2}}\left[ \frac{2\left( v_{e\parallel }v_{i\parallel
}-v_{i}^{2}\right) }{v_{r}^{2}}\left( 1+\frac{v_{r\parallel }^{2}}{v_{r}^{2}}%
\right) -\frac{v_{e\parallel }v_{r\parallel }v_{i\bot }^{2}}{v_{r}^{4}}%
\right] \Bigg\} \left[ \rho _{2}^{2}K_{1}(\rho _{2})\right] ^{2} \\
&+& \left[ \frac{v_{e\parallel }v_{i\parallel }-v_{i}^{2}}{v_{r}^{2}}\left( 
\frac{2v_{r\parallel }^{2}}{v_{r}^{2}}+\frac{v_{i\bot }^{4}}{4v_{r}^{4}}%
\right) +\frac{v_{r\parallel }v_{e\parallel }v_{i\bot }^{2}}{v_{r}^{4}}%
\left( \frac{v_{i\bot }^{2}}{2v_{r}^{2}}-2\right) \right] \left[ \rho
_{2}^{2}K_{2}(\rho _{2})\right] ^{2}.  \nonumber
\end{eqnarray}%
Here $\kappa _{2}^{2}=4/\delta ^{2}+1/\lambda ^{2}$, $\rho _{2}=\kappa _{2}s$%
. The cyclotron motion and the drift of the guiding center of the electron
are coupled to each other. Therefore the perturbation of the cyclotron
motion causes an additional perturbation of the guiding center motion. This
effect is given by the first term $\mathcal{Q}_{0}$ in Eq.~(\ref{eq:a73})
which depends on magnetic field through the cyclotron radius $a$ in the
prefactor, while the arguments of the modified Bessel functions in Eq.~(\ref%
{eq:a74}) do not depend on magnetic field. In the other terms $\mathcal{Q}%
_{1}$ and $\mathcal{Q}_{2}$ the arguments of the Bessel functions $\rho _{1}$
and $\rho _{2}$ correspond to the first and second cyclotron harmonic
perturbations, respectively.

For $v_{i\perp }=0$ we have $\mathcal{Q}_{0}=0$ and 
\begin{equation}
\label{eq:a77}
\mathcal{Q}_{1}=\frac{v_{i\parallel }}{v_{r\parallel }}\left\{ \rho _{1}^{4}%
\left[ K_{0}^{2}(\rho _{1})-K_{2}^{2}(\rho _{1})\right] +\left[ 2+\frac{6}{%
(\kappa _{1}\delta )^{2}}\right] \rho _{1}^{4}K_{1}^{2}(\rho _{1})-\frac{4}{%
(\kappa _{1}\delta )^{4}}\rho _{1}^{5}K_{0}(\rho _{1})K_{1}(\rho
_{1})\right\} ,
\end{equation}%
\begin{equation}
\label{eq:a78}
\mathcal{Q}_{2}=\frac{2v_{i\parallel }}{v_{r\parallel }}\left[ \rho
_{2}^{2}K_{2}(\rho _{2})\right] ^{2}.
\end{equation}%
With the help of the recursion relations of the modified Bessel functions it
is easy to see that the resulting energy transfer $\langle \Delta
E_{i}^{(2)}\rangle _{\mathrm{II}}$ agrees with the corresponding $\mathrm{O}%
(a^{2}/\delta ^{2})$-term of the energy transfer, Eq.~(\ref{eq:a59}), where
the limit of parallel ion motion $v_{i\perp }\rightarrow 0$ was taken before
the limit $a\ll \delta $.

\section{Discussion and Conclusion}
\label{sec:disc}

In this paper, we have presented a detailed theoretical investigation of the
energy transfer of a uniformly moving heavy ion due to the binary collision
(BC) with the magnetized electrons. The BC energy transfer can only be evaluated
explicitly in closed form in the limiting cases of a vanishing and an infinitely
strong magnetic field. The BC treatment developed here is valid for arbitrary
strengths of the magnetic field and arbitrary shapes of the interaction potential
up to second order in the interaction strength. The purpose of this work was to
investigate the ion energy transfer for finite magnetic fields which is explicitly
calculated for a regularized and screened potential which is both of finite range
and less singular than the Coulomb interaction at the origin and as the limiting
cases involves the Debye (i.e., screened) and Coulomb potentials. Two particular
cases have been considered in detail: (i) Ion motion parallel to the
magnetic field with an arbitrary strength. The energy transfer involves all harmonics
of the electron cyclotron motion. (ii) The ion arbitrary motion with respect to
the strong magnetic field when the electron cyclotron radius is much smaller than
other characteristic length scales (e.g., screening length, pitch of electron helix
etc.). We show that in the latter case the energy transfer receives two contributions
which are responsible for the electron guiding center and cyclotron orbit perturbations.

We would like to mention that our current results leave still some questions open.
It is clear that for the validity of the second--order perturbation BC theory
developed here more critical are the domains of the small relative velocities
$v_{r}$ and/or impact parameters $s$. Moreover, for the binary electron--ion
collisions in a magnetic field as given by the equation of motion (\ref{eq:a6})
there are less integrals of motion than degrees of freedom which indicates the
possibility of the chaotic dynamics in the system~\cite{gut90,sch00,bhu02}. This
immediately raises the question whether a perturbative treatment as proposed in
this paper can be applied at all. We will address this issue in the fortcoming
studies by showing some examples for the energy transfer obtained from a numerical
solution of the equation of motion~(\ref{eq:a6}) and by making some comparison with
the perturbative treatment. This topic is presently under investigation and will
be published elsewhere.

\begin{acknowledgments}
This work has been supported by the Armenian Ministry of Higher Education
and Science (Grant No.~0401) and The Armenian National Science and Education
Foundation (ANSEF) (Project~No.~PS87-01).
\end{acknowledgments}

\appendix

\section{Integrated energy transfer for parallel ion motion}
\label{sec:app1}

The integration of the energy transfer Eq.~(\ref{eq:a45}) with Eqs.~(\ref%
{eq:a46}) and (\ref{eq:a47}) with respect to the impact parameter $s$ is
facilitated by using the following relations for the Bessel functions 
\begin{equation}
\label{eq:ap1}
\int_{0}^{\infty }J_{n}\left( k_{\bot }s\right) J_{n}\left( k_{\bot
}^{\prime }s\right) sds=\int_{0}^{\infty }J_{n-1}\left( k_{\bot }s\right)
J_{n-1}\left( k_{\bot }^{\prime }s\right) sds=\frac{1}{k_{\bot }}\delta
\left( k_{\bot }^{\prime }-k_{\bot }\right) .
\end{equation}%
Using this relation we integrate the energy transfer Eq.~(\ref{eq:a45}) with
respect to the impact parameter. This yields 
\begin{eqnarray}
\label{eq:ap2}
\int_{0}^{\infty }\langle \Delta E_{i\parallel }^{(2)}\rangle sds &=&\left( 
\frac{Ze\!\!\!/^{2}}{\delta }\right) ^{2}\frac{4v_{i\parallel }}{%
mv_{r\parallel }^{3}}\sum\limits_{n=1}^{\infty }n^{2}\left\{ 3\Phi
_{n}\left( k_{\parallel },a\right) +k_{\parallel }\frac{\partial }{\partial
k_{\parallel }}\Phi _{n}\left( k_{\parallel },a\right) \right. \\
&+&\left. \frac{\delta ^{2}}{2n}\left[ \Psi _{n-1}\left( k_{\parallel
},a\right) -\Psi _{n+1}\left( k_{\parallel },a\right) \right] \right\}
_{k_{\parallel }=n/\delta },  \nonumber
\end{eqnarray}%
where
\begin{eqnarray}
\label{eq:ap3}
\Phi _{n}\left( k_{\parallel },a\right) &=&\int_{0}^{\infty }U_{n}\left(
k_{\parallel },a,s\right) sds=\frac{\left( 2\pi \right) ^{4}}{4}%
\int_{0}^{\infty }U^{2}\left( k_{\parallel },k_{\perp }\right)
J_{n}^{2}\left( k_{\bot }a\right) k_{\perp }dk_{\perp },   \\
\label{eq:ap4}
\Psi _{n}\left( k_{\parallel },a\right) &=&\int_{0}^{\infty }V_{n+1}\left(
k_{\parallel },s,a\right) sds=\int_{0}^{\infty }V_{n}\left( k_{\parallel
},a,s\right) sds   \\
&=&\frac{\left( 2\pi \right) ^{4}}{4}\int_{0}^{\infty }U^{2}\left(
k_{\parallel },k_{\perp }\right) J_{n}^{2}\left( k_{\bot }a\right) k_{\perp
}^{3}dk_{\perp }.  \nonumber
\end{eqnarray}%
Using the recurrent relations between the Bessel functions we obtain 
\begin{equation}
\label{eq:ap5}
\Psi _{n-1}\left( k_{\parallel },a\right) -\Psi _{n+1}\left( k_{\parallel
},a\right) =\frac{2n}{a}\frac{\partial }{\partial a}\Phi _{n}\left(
k_{\parallel },a\right) .
\end{equation}%
Thus the integrated energy transfer is expressed only by the functions $\Phi
_{n}$ 
\begin{eqnarray}
\label{eq:ap6}
\int_{0}^{\infty }\langle \Delta E_{i\parallel }^{(2)}\rangle sds &=&\left( 
\frac{Ze\!\!\!/^{2}}{\delta }\right) ^{2}\frac{4v_{i\parallel }}{%
mv_{r\parallel }^{3}}   \\
&\times & \sum_{n=1}^{\infty }n^{2}\left\{ 3\Phi _{n}\left( k_{\parallel
},a\right) +k_{\parallel }\frac{\partial }{\partial k_{\parallel }}\Phi
_{n}\left( k_{\parallel },a\right) +\frac{\delta ^{2}}{a}\frac{\partial }{%
\partial a}\Phi _{n}\left( k_{\parallel },a\right) \right\} _{k_{\parallel
}=n/\delta }.  \nonumber
\end{eqnarray}%
As an important particular case consider the regularized and screened
interaction potential Eq.~(\ref{eq:a48}). For this potential Eq.~(\ref%
{eq:ap3}) for the function $\Phi _{n}$ yields \cite{gra80} 
\begin{equation}
\label{eq:ap7}
\Phi _{n}\left( k_{\parallel },a\right) =\frac{2}{\lambda ^{-2}-d^{-2}}\left[
I_{n}\left( \kappa a\right) K_{n}\left( \kappa a\right) -I_{n}\left( \chi
a\right) K_{n}\left( \chi a\right) \right] -\frac{1}{2\kappa ^{2}}%
H_{n}\left( \kappa a\right) -\frac{1}{2\chi ^{2}}H_{n}\left( \chi a\right) ,
\end{equation}%
where $\kappa $, $\chi $ and $d$ have been introduced in Sec.~\ref{sec:s3}
and 
\begin{equation}
\label{eq:ap8}
H_{n}\left( \xi \right) =\xi \frac{\partial }{\partial \xi }\left[
I_{n}\left( \xi \right) K_{n}\left( \xi \right) \right] =\xi \left[
I_{n}^{\prime }\left( \xi \right) K_{n}\left( \xi \right) +I_{n}\left( \xi
\right) K_{n}^{\prime }\left( \xi \right) \right] .
\end{equation}%
Note that $H_{n}\left( \xi \right) =T_{n}\left( \xi ,\xi \right) $, where $%
T_{n}\left( x,y\right) $ is defined in Eq.~(\ref{eq:a51}). From Eq.~(\ref%
{eq:ap7}) one can derive the function $\Phi _{n}$ for screened and Coulomb
potentials. For the screened but unregularized potential (i.e. $\lambdabar
\rightarrow 0$) all terms in Eq.~(\ref{eq:ap7}) containing $\chi $ vanish
and the function $\Phi _{n}$ becomes 
\begin{equation}
\label{eq:ap9}
\Phi _{n}\left( k_{\parallel },a\right) =-\frac{1}{2\kappa ^{2}}H_{n}\left(
\kappa a\right) .
\end{equation}%
In a limit $\lambda \rightarrow \infty $, i.e. in the case of unscreened
Coulomb potential in Eq.~(\ref{eq:ap9}) the variable $\kappa $ is replaced
by $\left\vert k_{\parallel }\right\vert $.

\section{Integrated energy transfer for an infinitely strong magnetic field}
\label{sec:app2}

Consider the integrated energy transfer for an arbitrary ion motion and for
infinitely strong magnetic field. The integration of Eq.~(\ref{eq:a64}) with
respect to $s$ involves two integrals of the functions $T_{12}^{2}\left(
s\right) $ and $T_{01}\left( s\right) T_{03}\left( s\right) $ which can be
evaluated employing the relation~(\ref{eq:ap1}). In a general case with $%
q\neq 0$ we obtain 
\begin{equation}
\label{eq:ap10}
\mathcal{T}\left( q,\varkappa \right) =\int_{0}^{\infty }\mathcal{T}%
_{12}^{2}\left( q,s\right) sds=\int_{0}^{\infty }\mathcal{T}_{01}\left(
q,s\right) \mathcal{T}_{03}\left( q,s\right) sds=\frac{\left( 2\pi \right)
^{4}}{4}\int_{0}^{\infty }U^{2}\left( q,k_{\bot }\right) k_{\bot
}^{3}dk_{\bot }.
\end{equation}%
Obviously the integrations in Eq.~(\ref{eq:ap10}) require that the
interaction potential must decay faster than $r^{-1}$ at large distances and
must increase slower than $r^{-1}$ at small ones. In particular, for
regularized screened potential from Eq.~(\ref{eq:ap10}) we find 
\begin{equation}
\label{eq:ap11}
\mathcal{T}\left( q,\varkappa \right) =\frac{2q^{2}\lambda ^{2}+\varkappa
^{2}+1}{2\left( \varkappa ^{2}-1\right) }\ln \frac{q^{2}\lambda
^{2}+\varkappa ^{2}}{q^{2}\lambda ^{2}+1}-1
\end{equation}%
with $\varkappa =\lambda /d=1+\lambda /\lambdabar $ which at $q=0$ is
simplified to 
\begin{equation}
\label{eq:ap12}
\mathcal{U}_{0}(\varkappa ) =\mathcal{T}\left( 0,\varkappa
\right) =\frac{\varkappa ^{2}+1}{\varkappa ^{2}-1}\ln \varkappa -1.
\end{equation}%
Thus the $s$-integrated energy transfer in the presence of an infinitely
strong magnetic field reads 
\begin{equation}
\label{eq:ap13}
2\pi \int_{0}^{\infty }\langle \Delta E_{i}^{(2)}\rangle _{\mathrm{I1}}sds=%
\frac{2\pi Z^{2}e\!\!\!/^{4}v_{i\perp }^{2}}{mv_{r}^{6}}\mathcal{U}%
_{0}\left( \varkappa \right) \left( 2v_{e\parallel
}^{2}-v_{i}^{2}-v_{e\parallel }v_{i\parallel }\right) .
\end{equation}%
The quantity $\mathcal{U}_{0}\left( \varkappa \right) $ in Eq.~(\ref{eq:ap13}%
) can be treated as a modified Coulomb logarithm.

\end{document}